\documentclass[onecolumn,a4paper,showpacs,preprintnumbers,amsmath,amssymb,nofootinbib,floatfix]{revtex4}
\def \be {\begin{equation}}

\def \ee {\end{equation}}
\def \bea {\begin{eqnarray}}
\def \eea {\end{eqnarray}}
\usepackage{natbib}
\usepackage{graphicx}
\usepackage{dcolumn}
\usepackage{bm}
\usepackage{epsfig}
\usepackage{xcolor}

\begin{document}

\title{Forecasts analysis on varying-$\alpha$ theories from gravitational wave standard sirens}

\author{L. R. Cola\c{c}o}$^{1}$ \email{colacolrc@gmail.com}
\author{R. F. L. Holanda$^{1}$} \email{holandarfl@gmail.com}
\author{Rafael C. Nunes $^{2, 3}$} \email{costa.nunes@ufrgs.br}
\author{J. E. Gonzalez$^{4}$}
\email{javiergonzalezs@academico.ufs.br}

\affiliation{$^1$ Universidade Federal do Rio Grande do Norte, Departamento de F\'{i}sica Te\'{o}rica e Experimental, 59300-000, Natal - RN, Brasil.\\}
\affiliation{$^2$ Instituto de F\'{i}sica, Universidade Federal do Rio Grande do Sul, 91501-970 Porto Alegre RS, Brazil\\}
\affiliation{$^3$ Divis\~ao de Astrof\'isica, Instituto Nacional de Pesquisas Espaciais, Avenida dos Astronautas 1758, S\~ao Jos\'e dos Campos, 12227-010, SP, Brazil\\}
\affiliation{$^4$ Departamento de Física, Universidade Federal de Sergipe, Aracaju, SE 49100-000, Brasil}

\begin{abstract}

{ Motivated by future gravitational waves observations, we perform  forecasts analysis to constrain a possible time variation of the fine structure constant ($\alpha$) within the context of the so-called runaway dilaton model. For this purpose, some gravitational-wave standard sirens mock data within the perspective of Einstein Telescope and LISA mission were considered jointly with current strong gravitational lensing systems observations. We find that future standard sirens observations can also play an important role in the search for possible variations of $\alpha$ within the methodology presented in this work.}

\end{abstract}
\maketitle

\section{Introduction}

The possibility that the fundamental constants of nature are actually not constants, but time-evolving quantities following the slow pace of the cosmological evolution, had a long discussion in the literature. This question was probably first discussed by Dirac \cite{Dirac1} in his ``large numbers hypothesis'' and by Milne, \cite{Milne1935} introducing a possibility for a time variation of the gravitational coupling $G$. Jordan speculated that the fine structure constant $\alpha$ together with $G$ could be both space and time-dependent \cite{jordan1937,jordan1939}  (see also \cite{2020JCAP...07..060B,2012PhR...513....1C} and references therein). Later on, Brans and Dicke also proposed a time variation possibility on $G$, driven by a dynamical scalar field coupled to curvature \cite{PhysRev.124.925}, while Gamow also triggered subsequent speculations on the possible variation of the fine structure constant \cite{PhysRevLett.19.759}. The possibility that the particle masses and the speed of light could also drift with the cosmic evolution also have been proposed and discussed \cite{1999PhRvD..59d3516A,2005AmJPh..73..240E,2018ApJ...867...50C,2017JCAP...02..012C,2021JCAP...11..034M,2021MNRAS.506.2181L}. Several other pioneers theoretical motivations have been introduced that lead to temporal variation of the constants of nature (see \cite{Uzan2011,Martins2017} for a review related to theoretical and experimental research on the variation of the fundamental constants).

{ Regardless of the merit of such approaches, one thing is clear: one detection of varying fundamental constants of nature would be revolutionary. For instance, a possible fundamental couplings  dynamical violates the Einstein Equivalence Principle, and as consequence, gravity could be a not purely geometric phenomenon. Then, gravity laws should be modified at large and/or small scales. Also, could show that there is a fifth force of nature waiting out there to be discovered \cite{Damour:2002vu}. Variations of the fine structure constant could imply a non-conservation of the photon number along
geodesic. Such a non-conservation can have several observational consequences, such as, 
that the CMB radiation does not obey the adiabaticity, and, then,  the CMB would not be an blackbody radiation. The peak luminosities of SNe Ia also depend on a possible time variation of Newton’s constant and $\alpha$. Thanks to current technological advances, several observational data enable us to probe such approaches. However, even improved null results are important and indeed extremely useful as technology improves itself. That can be perceived by noting that the natural scale for the cosmological evolution of any constant of nature, if driven by a fundamental scalar field, would be the Hubble time. Therefore, we would expect a relative Hubble time drift rate of the order of $10^{-10}$ per year. However, current local bounds from laboratory atomic clock comparison experiments are about six orders of magnitude stronger yet \cite{Rosenband}. Thus dynamical scalar fields must be `slow-rolling', something that has analogies with dark energy and inflation.}

On the other hand, from a more modern perspective, the observational evidence that our Universe is currently in a stage of accelerated expansion leads us to introduce some extra degrees of freedom compared to general relativity (GR), as dark energy models and modified gravity theories, of which also predict cosmic time variation of the fundamental constants. These include scalar-tensor theories \cite{2003sttg.book.....F,2016CQGra..33iLT01N,hees,Bruck2015}, modified Teleparallel gravity \cite{Nunes2017,Said2020}, running vacuum models \cite{Nunes2017b}, Bekenstein-Sandvik-Barrow-Magueijo theory \cite{1982PhRvD..25.1527B,2012PhRvD..85b3514B}, extra-dimensions \cite{1997PhR...283..303O}, dynamical dark energy models \cite{2016PhRvD..93b3506M,2015PhRvD..91j3501M,2017PhRvC..96d5802M,2021ApJ...922...19L}. Briefly, in the astronomical context, tight constraints on $\Delta \alpha/\alpha$ are obtained from white dwarf observations. For instance, by using white dwarf gravitational potential, the Refs. \cite{Landau:2020vkr,Bainbridge:2017lsj} put limits on $\Delta \alpha/\alpha$ at the level $(2.7 \pm 9.1) \times 10^{-5}$. Very recently, from observations up to $z \approx 7.1$4 (by the so-called many-multiplet method), new quasars spectral observations show no evidence for a redshift variation of the fine-structure constant  \cite{Wilczynska}. On the other hand, when these new measurements were combined with a large existing sample at lower redshifts,  a spatial variation of $\alpha$ was preferred over a no-variation model at the $3.7 \sigma$ level (see other discussions about possible $\alpha$ spatial variation in \cite{webb1999,Ubachs:2017zmg,2021MNRAS.500....1M}). However, it is important to comment that the authors of the Ref. \cite{Lee} showed to exist a degeneracy between the absorption structure and turbulent models in quasar analyses, each giving different $\Delta \alpha / \alpha$ values. Naturally, this fact adds a substantial additional random uncertainty to $\Delta \alpha /  \alpha$ \cite{Wilczynska}. 

Other tests for a possible $\alpha$ variation have been performed by using distinct astrophysical observables, such as galaxy cluster \cite{galli},  Big-Bang Nucleosynthesis \cite{BBN}, black hole in a high gravitational potential \cite{Hees:2020gda}, among others\cite{Milakovic:2020tvq,Kraiselburd:2018uac}. Variation of $\alpha$ and $m_e$  straightly modify the recombination history at $z \approx 1100$,  changing the temperature and polarization anisotropies of the cosmic microwave background (see  Ref. \cite{Hart:2019dxi}). In this line, tight constraint on $\Delta \alpha / \alpha$ was performed by the Planck satellite  data \cite{Planck2015,2018MNRAS.474.1850H}, $\Delta \alpha / \alpha \approx 10^{-3}$. However, it is worth commenting that such a limit is inferred for a specific cosmological model, namely: the flat $\Lambda$CDM model, with purely adiabatic initial conditions and an almost scale-invariant power spectrum, being weakened if the parameter space is allowed to vary in the number of relativistic species or the helium abundance (see Figs. 5 and 6 in \cite{Planck2015}). As local methods, atomic clock measurements \cite{Hinkley2013}, spectroscopy of radio-frequency transitions \cite{2013PhRvL.111f0801L}  and isotope ratio measurements \cite{Dijck:2020kfb} have also been used to obtain the tightest limits on  $\Delta \alpha / \alpha $.

From a theoretical point of view, Ref. \cite{hees} showed that the Einstein equivalence principle in the electromagnetic sector is violated in a general class of modified gravity theories that have a non-minimal multiplicative coupling between the usual electromagnetic part of matter fields and a new scalar field. In such a framework the entire electromagnetic sector of the theory is affected, leading to $\alpha$ variation. In this line, a particular class of string theory inspired-models that produce a temporal variation of alpha is the so-called runaway dilaton model \cite{damour1,Martins2019,Martins2018,Martins2015} (string theories at low energy predict the existence of dilaton, a scalar partner of spin-2 graviton). In order to check a possible temporal variation of the fine-structure constant in runaway dilaton scenarios, some methods using astronomical data have been developed in recent years\cite{Kamal2021,leo1,Colaco2019,Holanda2016JCAP,Holanda2016JCAP2,Martins2015,Martins20152}.   
\\

{   On the other hand, from an observational perspective, looking for new astrophysical sources, through a direct manifestation of
gravitational effects, can provide rich physical information about the nature of gravity and/or the dark sector of the Universe (dark energy and dark matter). In that regard, the gravitational wave (GW) astronomy provides an unprecedented opportunity to test physics in that direction. Currently, more than 100 coalescing compact binary events have already been observed during the three running stages of the LIGO/VIRGO mission \cite{theligoscientificcollaboration2021gwtc3}. One of the most promising prospects is the observation of standard siren (SS) events \cite{1986Natur.323..310S,Holz_2005}. The latter are the GW analog of the astronomical standard
candles and might be a powerful tool in view of constraining cosmological parameters through the information encoded in the luminosity distance provided by these events. To date, one event has been observed through a binary neutron star (BNS) merger at z = 0.01, namely the GW170817 event \cite{Abbott_2017}. Preliminary cosmological information and the consequences of this observation are important to the understanding of our Universe locally. These observations were used to measure the Hubble constant \cite{LIGOScientific:2017adf} and also to impose strong constraints on modified gravity/cosmology theories (see \cite{Kase:2018aps} for a review).}

{  The detectability rate of the SS events from the current LIGO/VIRGO sensitivity is expected to be very low, as well as difficult to reach large
cosmic distances. The central importance of GW astronomy is testified by the plans for the construction of several GW observatories in the future, such as the underground-based interferometers ET \cite{Maggiore:2019uih} and Cosmic Explore \cite{Reitze:2019iox}, and space-based interferometers such as LISA \cite{LISA:2017pwj}, DECIGO \cite{Kawamura:2020pcg}, and TianQin \cite{TianQin:2015yph}, among others, to observe GWs in the most diverse frequency bands. The implications of cosmological studies using the SS have motivated focused studies on the nature of dark energy, modified gravity, dark matter, and several other fundamental questions in modern cosmology \cite{Cai:2016sby,Du:2018tia,Zhang:2018byx,Yang:2019vni,Fu:2019oll,Cai:2017aea,Allahyari:2021enz,Belgacem:2017ihm,DAgostino:2019hvh,Nishizawa:2019rra,Bonilla:2019mbm,Odintsov:2022cbm,Cai:2021ooo,Matos:2021qne,Jiang:2021mpd,Pan:2021tpk,Tasinato:2021wol,Bonilla:2021dql,Mukherjee:2020mha,Kalomenopoulos:2020klp,Baker:2020apq,Mastrogiovanni:2020gua,Belgacem:2019zzu,Nunes:2019bjq,Harry:2022zey,Ezquiaga:2021ler}. It is important to emphasize that GWs observations are not affected by well-known systematical effects related to supernova science.
For instance, are not affected by propagation effects, dust extinction, or microlensing by stars \cite{2021MNRAS.503.3326C,2021arXiv211207635Y,2023arXiv230401202L}. Then, GWs observations science can be considered attractive from the perspective of cosmological tests.}

This work is divided into two parts. {Firstly,  we revisit  the Ref.\cite{leo1}, where geometrical measurements of strong gravitational lensing systems (SGL)  and Type Ia supernovae (SNe Ia) sample were used  to constrain a possible time evolution of the fine-structure constant induced for the well-known runaway dilaton model. Particularly, we redo the analysis of the \cite{leo1} in a more robust way by taking into account the correlation between SNe data in analysis. In a second part, as a step forward concerning the Ref. \cite{leo1}, we will perform  forecasts analysis based on the generation of some standard sirens catalogs within the perspective of two future gravitational waves observatories, namely, Einstein Telescope and LISA,  jointly with current SGL observations. We obtain that the results coming from GWs mock data (+SGL) analysis are competitive with the current ones from SNe Ia (+SGL). It is very worth to comment that the method proposed by \cite{leo1} is  independent of the  $M_B$,  which is used to fit Type Ia supernovae sample. As it is largely known,  this parameter is at strong statistical tension in $\Lambda$CDM framework, and may be the cause of the $H_0$ tension \cite{efstathiou2021h0,Camarena_2021,Nunes2021c} and being able to add new correlations on models beyond the $\Lambda$CDM paradigm.} This paper is structured as follows. Next Section we present the methodology. In Sections \ref{model} and \ref{data}, we present our theoretical framework and the data sets used in this work, respectively. In Section \ref{results}, we discuss the main results of our analysis. In Section \ref{final}, we outline our final considerations and perspectives.

\section{Methodology}
\label{Methodology}

In what follows, we describe our methodology. The Strong Gravitational Lensing (SGL) effect is one prediction of GR occurring when the source ($s$), lens ($l$), and the observer ($o$) are at the same line-of-sight to form the Einstein ring, a ring-like structure with angular radius $\theta_E$ \cite{cao2015}. It is a purely gravitational phenomenon where, in the cosmological scenario, a lens can be a foreground galaxy or galaxy cluster placed between the source and observer. Under the assumption of the singular isothermal sphere (SIS) model, the lens mass distribution, $\theta_E$, is given by \cite{cao2015,Refsdal}: 

\begin{equation}
\label{theta_E}
    \theta_E = 4\pi \frac{D_{A_{ls}}}{D_{A_{s}}} \frac{\sigma_{SIS}^{2}}{c^2},
\end{equation}
where $D_{A_{ls}}$ is the angular diameter distance from the lens to the source, $D_{A_{s}}$ is the angular diameter distance to the source, $c$ is the speed of light (SoL), and $\sigma_{SIS}$ is the velocity dispersion measured under SIS model assumption. 

On galaxy scales, such a phenomenon has been largely used to ascertain gravitational and cosmological theories and fundamental physics. Particularly, SGL systems observed and detected by SLACS, LSD, SLS2, and BELLS surveys had significant progress in the last years {  due to an increase in the accuracy of lens-modeling and to precise limit obtaining on distinct cosmological parameters} \cite{Suyu,H0LiCOW,Shajib,Birrer,leo2}. For instance, \cite{leo1} provided a robust test based on SGL and type Ia Supernovae observations in order to put new bounds on a possible time variation of the fine-structure constant ($\alpha$). This method is based on Eq. (\ref{theta_E}) for lenses and the observational quantity $D$ defined by:

\begin{equation}
\label{eq2}
D \equiv \frac{D_{A_{ls}}}{D_{A_s}} = \frac{ \theta_E c_s^2}{4\pi \sigma_{SIS}^{2}},
\end{equation}
where $c_s$ is the SoL measured at $z_s$. In fact, \cite{leo1} extended the original method provided by \cite{holg} which investigates any deviation of the Cosmic Duality Distance Relation (CDDR) through SGL and SNe Ia observations. According to the definition of the fine-structure constant ($\alpha_s=e^2/\hbar c_s$), the Eq. (\ref{eq2}) can be rewritten as:

\begin{equation}
    D\equiv \frac{D_{A_{ls}}}{D_{A_{s}}}=\frac{e^4\theta_E}{\hbar^2 \alpha_s^2 4\pi \sigma_{SIS}^{2}}.
\end{equation}
{  Although this equation introduces the key quantities for our paper, a more general and appropriate approach will be used. This occurs  due to recent studies using SGL systems have shown that the SIS model may not be an accurate representation of the lens mass distribution \cite{Koopmans,Auger,Barnab,Sonnenfeld,cao2015,holanda2017,YCHEN}. Following several recent works, a power law ($\rho \propto r^{-\Gamma}$) for the lens mass distribution will be used (see Ref. \cite{cao2015} for mathematical details)}.

From another perspective, under a flat Universe assumption with comoving distance between the lens and the observer being $r_{ls} = r_s-r_l$, and using the relations $r_s = (1 + z_s) D_{A_s}$, $r_l = (1 + z_l) D_{A_l}$, $r_{ls} = (1 + z_s) D_{A_{ls}}$, it is possible to obtain \cite{holg}:

\begin{equation}
D = 1-\frac{(1+z_l)}{(1+z_s)}\frac{D_{A_l}}{D_{A_s}}.
\end{equation}
Considering a possible deviation of CDDR by $D_{A_i}=D_{L_i}\eta^{-1}(z_i)(1+z_i)^{-2}$, we can obtain:

\begin{equation}
D = 1- \frac{(1+z_s)D_{L_l}}{(1+z_l)D_{L_s}}\frac{\eta(z_s)}{\eta(z_l)},
\end{equation}
where $D_{L_l}$ and $D_{L_s}$ are the luminosity distances to lens and source, respectively, and $\eta(z_i)$ captures any deviation of CDDR.

\section{Varying-$\alpha$ in Scalar-Tensor Gravity}
\label{model}

Modified gravity theories related to a non-minimal multiplicative coupling between an extra scalar field and the usual matter Lagrangian lead to violations of the Einstein Equivalence Principle (EEP) in the electromagnetic sector \cite{hees,hees2}. In this context, the usual matter Lagrangian is given by \cite{hees}:

\begin{equation}
    S_{\mathrm{mat.}} = \sum_i \int d^4x \sqrt{-g} h_i(\phi) \mathcal{L}_{i}(g_{\mu \nu}, \Psi_i),
\end{equation}
where $\mathcal{L}_i$ are the Lagrangians for the different matter fields ($\Psi_i$), and $h(\phi)$ is a function of the scalar field. In the electromagnetic sector, the fine-structure constant $\alpha$ and the CDDR change over cosmological time\footnote{In this type of theory, a variation of $\alpha$ can arise from a varying $\mu_0$ (vacuum permeability) or from a variation of charge of the elementary particles. Both interpretations lead to the same modified expression of $\alpha$ \cite{Uzan:2010pm,Observables}.}, and both changes are intimately and unequivocally related by \cite{hees}:

\begin{eqnarray}
\label{eq7}
    \frac{\Delta \alpha}{\alpha} && \equiv \frac{\alpha(z)-\alpha_0}{\alpha_0} = \frac{h(\phi_0)}{h(\phi)}-1 = \eta^2(z) - 1
\end{eqnarray}

In this work, we will focus on the so-called Runaway Dilaton Model, a string theory-inspired model capable of delineating a time-variation of $\alpha$ close to the present day \cite{damour1,Damour12,Martins2019,Martins2018,Martins2015}. Such a model is a particular case of scalar-tensor theories of gravity inspired by a multiplicative coupling between an extra scalar-field and the usual matter Lagrangian \cite{Martins2015}. Basically, the model explores the string-loop modification of the four-dimensional effective low-energy action. {  Within the runaway of dilaton context, the time variation of $\alpha$ is given by}

\begin{equation}
 \frac{\Delta \alpha}{\alpha} (z) = \frac{1}{40} \beta_{had,0} \left[  1-e^{-(\phi(z)-\phi_0)}   \right].   
\end{equation}
{  Thus the behavior of $\Delta \alpha/\alpha$ close to the present day depends both on $\beta_{had,0}$ and the speed of the field $\phi_{0}^{'}$. Since we shall be interested in the evolution of $\alpha$ at relatively low redshifts, one could think of linearizing the field evolution by \cite{Martins2015}:}

\begin{equation}
    \phi \sim + \phi_0 + \ln{(a)},
\end{equation}
{  where $a$ is the scale factor. Thus, the equation (8) takes a simpler form by}

\begin{eqnarray}
\label{eq8}
    \frac{\Delta \alpha}{\alpha} && \approx -\frac{1}{40}\beta_{\mathrm{had,0}}\phi_{0}^{'}\ln{(1+z)} \equiv - \gamma \ln{(1+z)},
\end{eqnarray}
where $\gamma \equiv \frac{1}{40}\beta_{\mathrm{had,0}} \phi_{0}^{'}$, $\beta_{\mathrm{had,0}}$ is the current coupling value between dilaton and hadronic matter\footnote{The relevant parameter of the model is the coupling between dilaton and hadronic matter. The central hypothesis of the model is that all gauge fields couple to the same gauge coupling function ($B_F(\phi)$) \cite{Martins2015}.}, and $\phi_{0}^{'} \equiv \frac{\partial \phi}{\partial \ln{a}}$. {  It is important to stress that Eq.(\ref{eq8}) can be considered in low and intermediate redshifts. As shown in the second panel of Fig.1 of \cite{Martins20152}, the approach given by Eq.(\ref{eq8}) can still be of order unity by $z \approx 5$ for the values of the coupling that saturate the current bounds. Thus, the evolution of $\alpha$ can be calculated using the full equations.} 

Therefore, assuming $\alpha$ evolves like $\alpha(z)=\alpha_0\phi(z)$, where $\alpha_0$ is the current value of the fine-structure constant\footnote{The universal coupling of the dilaton to matter has a suggestive simplicity; a generalization for the universal $e^{-2\phi}$ coupling arises at the string tree level \cite{Damour:1994zq}.}
, and $\phi(z)$ is a scalar field that controls a time-variation of $\alpha$, the Eq. (\ref{eq7}) gives $\phi(z)=\eta^2(z)$. Thus, the equations (3) and (5) might be written, respectively, by:

\begin{equation}
     D=\frac{e^4\theta_E}{4\pi \alpha_{0}^{2} \hbar^2 \sigma_{SIS}^{2}}\phi^{-2}(z_s) = D_0\phi^{-2}(z_s)
\end{equation}
and
\begin{equation}
D =  1- \frac{(1+z_s)D_{L_l}}{(1+z_l)D_{L_s}}\frac{\phi^{1/2}(z_s)}{\phi^{1/2}(z_l)},
\end{equation}
where $D_0 \equiv e^4\theta_E /4 \pi\alpha_{0}^{2}\hbar^2 \sigma_{SIS}^{2}$. If $\Delta \alpha/\alpha = 0$, thus $\phi(z)=1$ and $D=D_0$. Combining Eq.s (11) and (12), it is possible to obtain:

\begin{equation}
    D_0 = \phi^2 (z_s) \left[ 1- \frac{(1+z_s)D_{L_l}}{(1+z_l)D_{L_s}}\frac{\phi^{1/2}(z_s)}{\phi^{1/2}(z_l)} \right].
\end{equation}
This is the equation we shall use to compare the model predictions with SGL, SNe Ia and GW observations.

\section{Data Set}
\label{data}

In this work, we desire to constrain possible departure from EEP described above using observational data obtained by probes which map the expansion history of the late-time universe (and in particular lying in the region $z < 3$). Our analysis is based on the Type Ia Supernovae distance moduli measurements from the Pantheon sample, strong gravitational lensing systems, and some mock data from gravitational wave standard sirens. In the following subsections, we present the different data sets used in our analysis. {  As commented earlier, the analysis by using SNe Ia and SGL was performed by the Ref.\cite{leo1}. We redo it by tanking into account the correlation between SNe data in analysis and  directly compare the new results with those coming from  GWs mock data (+SGL) forecasts.}

\subsection{Type Ia Supernovae}

{ We consider and concentrate the analyses of the present paper on the so-called Pantheon sample \cite{pantheon}, a widely refined sample of SNe Ia consisting of 1048 spectroscopically confirmed SNe Ia covering a redshift range of $0.01 \leq z \leq 2.3$. However, there is a Pantheon successor sample, the so-called Pantheon+ \cite{Brout:2022vxf}, which is not considered in this paper. Such a sample was built on the analysis framework of the original sample to combine an even larger number of SNe Ia and include the ones that are in galaxies with measured Cepheid distances. The new sample features an increased sample size and $z$ range, and better treatment of systematic uncertainties in comparison to the original Pantheon analysis.

{We transform the  apparent magnitude ($m_b$) sample into a $D_L$ dataset by using the relation: 

\begin{equation}
\label{MB_eq}
D_L=10^{(m_b-M_b-25)/5} \text{Mpc}, 
\end{equation}
where $M_b$ is an assumed absolute magnitude value, which is the same for all SNe data. {  As it will be shown, the specific $M_b$ is not relevant in our analyses. For this reason, we do not propagate the $M_b$ uncertainty into $D_L$.} The Pantheon data includes the covariance matrix of the $m_b$ of each SNe, which has to be also transformed into a  $D_L$-covariance matrix considering the transformation relation\footnote{The variables in bold correspond to vector representations of each data set.}:

\begin{equation}
cov(\bm D_{L}, \bm D_{L})=\left(\frac{\partial \bm D_{L}}{\partial \bm m_b}\right) cov(\bm m_b, \bm m_b)\left(\frac{\partial \bm D_{L}}{\partial \bm m_b}\right)^T,
\end{equation}
where $\left(\frac{\partial \bm D_{L}}{\partial \bm m_b}\right)$ represents the Jacobian matrix of the tranformation.} {  Within our methodology, we must use the luminosity distance from SNe Ia at the same (or approximately) SGL $z_l$ and $z_s$. Thus, we make a selection of SNe Ia for each lensing system according to the criteria: $| z_s-z_{SNe}| \leq 0.005$ and $| z_l-z_{SNe}| \leq 0.005$. Some systems have more than one couple of SNe Ia obeying the criteria for both redshifts, and others have none. Then we perform the weighted average for each system  based on SNe Ia samples obeying the selection criterion. In order to reduce uncertainties and taking into account the correlation between SNe data, we considered the covariance matrix $cov(\bm D_{L}, \bm D_{L})$ \cite{Schmelling_1995}} by:

\begin{equation}
\label{DLave}
\bar{D}_L = \frac{\sum_{i,j} {D_L}_iw_{ij}}{\sum_{i,j} w_{ij}},
\end{equation}
being the weight coefficients $w_{ij}=(cov^{-1}(\bm D_{L}, \bm D_{L}))_{ij}$.

The weighted averaged luminosity distances are correlated due to the intrinsic correlation in the SNe data and because, in some cases, there are some SNe that are used to calculate different $\bar{D}_L$. It can be shown that the covariance matrix of the averaged luminosity distances satisfies the relation:

\begin{equation}
\label{covDL}
cov( \bar D_{L_i},  \bar D_{L_j})= \sum_\alpha^{n_i} \sum_\gamma^{n_j}\frac{cov( D_{L_\alpha},  D_{L_\gamma})\left[\sum_\beta^{n_i}w_{\alpha \beta}\right]\left[\sum_\sigma^{n_j}w_{\gamma \sigma}\right]}{\left[\sum_\sigma^{n_i}\sum_\beta^{n_i}w_{\sigma \beta}\right]\left[\sum_\sigma^{n_j}\sum_\beta^{n_j}w_{\sigma \beta}\right]},
\end{equation}
where $n_i$ and $n_j$ correspond to the number of SNe considered to calculate the $i$-th and $j$-th averaged luminosity distances, respectively. A single SNe Ia is enough.

As well discussed in the literature, the peak luminosities of SNe Ia may depend on the time variation of Newton’s constant and the coupling constant $\alpha$. Considering a variation of these constants directly translates into a different peak bolometric magnitude, i.e, the distance modulus is modified \cite{Chiba,Kraiselburd}. From first principles, a varying-$\alpha$ framework must also change the rate of expansion of the Universe and then modify the prediction on the distance modulus. At the zero order, i.e., taking only changes on the background dynamics, these two effects must be the leading corrections to be taken into account for any specific model-dependent analysis at  the background level using geometrical measurements.

As explained earlier, in the methodology used here, our aim will be probing a varying-$\alpha$  possibility minimally independently of background dynamics corrections that these scenarios can predict. But, on the other hand, a specific scalar-tensor scenario is still required to induce the CDDR change on the cosmic time. Thus, the methodology here provides a test for the runaway dilaton field at the geometrical level regardless of the details that the dilaton field can induce on the $H(z)$ expansion and the $m_b$ module. Evidently, it does not cover more general scalar-tensor theories framework as Horndeski theory and its generalization, as well as many other perspectives in the scalar-tensor context. Therefore, our approach can be seen as a specific test to probe some specific scalar-tensor models with minimal assumptions on the background level.

\begin{figure*}
\label{DL_sample}
    \includegraphics[scale=0.52]{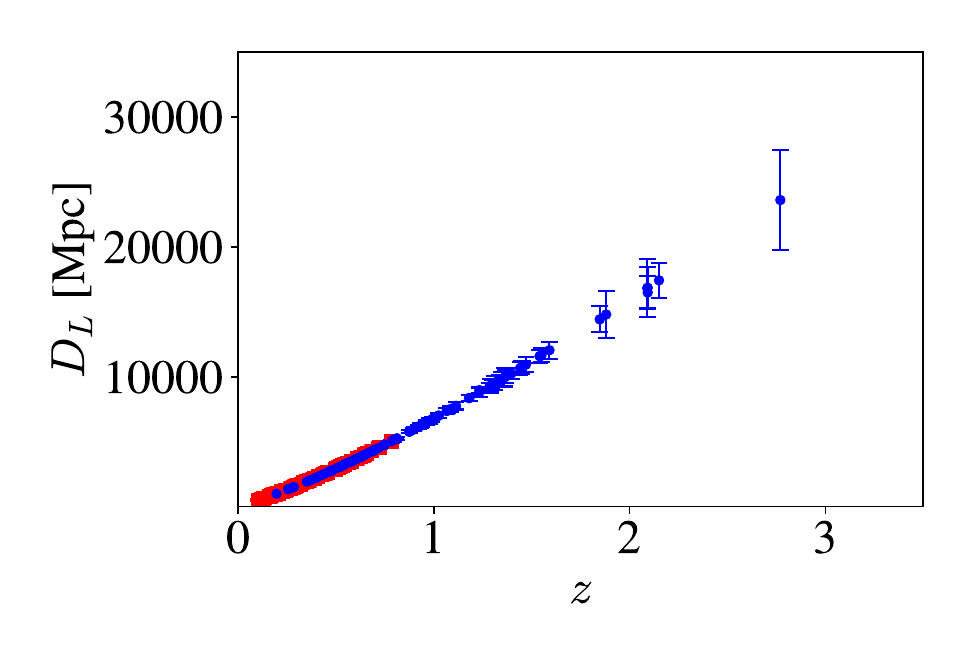}
    \includegraphics[scale=0.52]{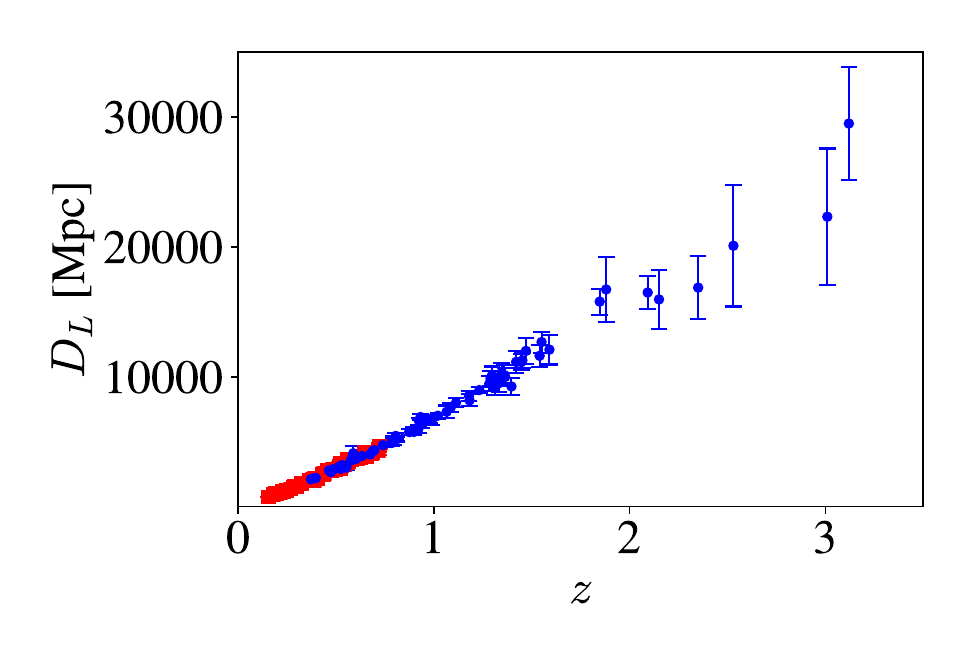}
    \caption{Left panel: Luminosity distance data obtained from ET mock data sample 1 (total of 97 points) together with LISA (total of 2 points). The blue data correspond to the luminosity distances from the observer to the source, and the red ones from observer to lens for each SGL system used in our analyses. Right panel: Same as in the left panel, but from ET mock data sample 2 (total of 69 points) together with LISA (total of 2 points).
}
\end{figure*}

\subsection{Gravitational Wave Standard Sirens}

Let us generate some mock data inspired by the possibility of future observation of standard siren (SS) events. We will create some SS mock catalogs, the GW analog of the astronomical standard candles, which can provide powerful information about the dynamics of the Universe up to high $z$.

For a given a GW strain signal, $h(t) = A(t) \cos [\Phi(t)]$, one can use the stationary-phase approximation for the orbital phase of the inspiraling binary system to obtain its Fourier transform $\tilde{h}(f)$. In this case, the  waveform of a coalescing binary system will take the form
\begin{equation}
\label{waveform}
\tilde{h}(f) = Q \mathcal{A} f^{-7/6} e^{i\Phi(f)}\ ,
\end{equation}
where $\mathcal{A} \propto 1/d_L$ is the luminosity distance to the redshift of the merger, and $\Phi(f)$ is the inspiral phase of the binary system. More details on the post-Newtonian coefficients and waveforms can be found in \cite{Rocco_Nunes_2019} (see references therein and Appendix A). Once the GW signal is defined, for a high enough signal-to-noise ratio (SNR), we can obtain upper bounds on the free parameters of the GW signal by means of the Fisher information analysis. Estimating $D_L(z)$ from GW standard sirens mock data is a well-consolidated methodology, and we refer to \cite{Rocco_Nunes_2019} and the references therein for a detailed description. In what follows, we briefly describe our methodology that is used to generate the SS mock catalog from the perspective of two future observatories, namely  Einstein Telescope (ET) and LISA.

The ET is a third-generation ground detector, covering the frequency range $1-10^4$ Hz. The ET is expected to be ten times more sensitive than the current advanced ground-based detectors. See \cite{ET_2020} for a presentation of the scientific objectives of the ET observatory. The ET conceptual design study predicts an order of $10^3-10^7$ BNS detection per year. Nevertheless, only a small fraction ($\sim 10^{-3}$) of them is expected to be accompanied by a short { Gamma-ray} burst observation. Assuming a detection rate of $\mathcal{O}(10^5)$, the events with short { Gamma-ray} bursts will be $\mathcal{O}(10^2)$ per year. In our simulations, we have considered two different samples. i) A catalog with 1000 BNS mock GW SS merger events up to $z = 2$. In this case, we perform a random sampling of the masses from a uniform distribution between [1-2]$M_\odot$. Let us {  refer} this sample by ET - mock data 1. ii) Let us consider a mass distribution of the astrophysical objects NS and BH with a random sampling of their masses from uniform distributions [1-2]$M_\odot$ and [3-10]$M_\odot$, respectively. This catalog also contains 1000 mock GW SS merger events, but with distributed events up to $z = 5$. Let us {\  denote} this sample by ET - mock data 2. For each event, we have estimated the measurement error on the luminosity distance by applying the Fisher matrix analysis. We calculated the SNR of each event and confirmed that it is a GW detection if SNR $>$ 8. Details of this methodology are well described in previous works \cite{ET_2011,ET_2017,Rocco_Nunes_2019}.

The LISA is a space-borne detector with a sensitivity peak of around 1 millihertz. Among astrophysical sources, LISA can reach include Galactic binaries, stellar-origin black hole binaries, and extreme-mass-ratio inspirals. LISA will also observe massive black hole binaries (MBHBs) from $10^4$ to $10^7$ solar masses. See \cite{amaroseoane2017laser} for a review of the scientific details of the LISA mission. The high SNR of the detected signals will allow for more precise parameter estimations. Among the most probable LISA sources with electromagnetic counterparts are MBHBs. In particular, MBHBs are supposed to merge in gas-rich environments and within the LISA frequency band allowing for electromagnetic followups to determine their redshift. The prospect of MBHBs that could have EM counterparts extends up to $z$ $\sim$7 providing a unique probe of the universe at high $z$. In this work, our catalog is based on the model presented in \cite{Tamanini2016}, where a semi-analytic framework allows tracing the galactic baryonic structures and dark matter mergers. Also, this methodology integrates the BH seeding at high $z$ and the delays between the merger of two galaxies and that of the massive BHs residing in the galaxies. For the purposes of the methodology of this work, we only take the category of population models named Pop III. {  Again, within our methodology, we must use the luminosity distance from GWs (simulated samples) in the same (or approximately) SGL $z_l$ and $z_s$. Thus, we make a selection of GWs such as:}
\begin{equation}
\bar{D}_L = \frac{\sum_i {D_L}_i/\sigma_{{D_L}_i}^{2}}{\sum_i 1/\sigma_{{D_L}_i}^{2}}, 
\label{eq:D_L-bar}
\end{equation}
\begin{equation}
\sigma_{\bar{D}_L}^2 = \frac{1}{\sum_i 1/ \sigma_{{D_L}_i}^{2}}.
\end{equation}

{As commented earlier, from the perspective of cosmological tests, gravitational waves are  considered attractive because they not be affected by, for instance, by propagation effects, dust extinction, or microlensing by stars \cite{2021MNRAS.503.3326C,2021arXiv211207635Y,2023arXiv230401202L}. }

\subsection{Strong Gravitational Lensing Systems}

In this section, we present a specific catalog containing 158 confirmed sources of SGL that will be used to perform the corresponding statistical analyses. Such catalog includes 118 SGL systems identical to the compilation of \cite{cao2015} obtained from SLOAN Lens ACS, BOSS Emission-line Lens Survey (BELLS), and Strong Legacy Survey SL2S, plus 40 new systems recently discovered by SLACS and pre-selected by \cite{Shu2017} (see Table I in \cite{2018MNRAS.478.5104L}). 

However, {  as commented earlier,  different studies using SGL systems have shown that the slopes of density profiles for individual galaxies exhibit a non-negligible deviation from the SIS model, indicating an inaccurate representation for the lens mass distribution \cite{Koopmans,Auger,Barnab,Sonnenfeld,cao2015,holanda2017,YCHEN}. Therefore, as done in the more recent works, we assume the power-law model (PLAW) for the mass distribution of lensing systems. Such a model basically assumes a spherically symmetric mass distribution with a more general power-law index $\Upsilon$ like $\rho \propto r^{-\Upsilon}$, where $\rho$ is the total mass distribution and $r$ is the spherical radius from the lensing galaxy center.} Assuming that the velocity anisotropy can be ignored and solving the spherical Jeans equation, it is possible to rescale the dynamical mass inside the aperture of size $\theta_{ap}$ projected to the lens plane and obtain

\begin{equation}
    \theta_E = \frac{4\pi \sigma_{ap}^{2}}{c^2}\frac{D_{A_{ls}}}{D_{A_s}} \Bigg(    \frac{\theta_{E}}{\theta_{ap}} \Bigg)^{2-\Upsilon} f(\Upsilon),
\end{equation}
where $\sigma_{ap}$ is the stellar velocity dispersion inside the aperture $\theta_{ap}$, and

\begin{eqnarray}
    f(\Upsilon) & \equiv & -\frac{1}{\sqrt{\pi}} \frac{(5-2\Upsilon)(1-\Upsilon)}{3-\Upsilon}\frac{\Gamma (\Upsilon-1)}{\Gamma (\Upsilon -3/2)} \nonumber \\
     && \times \left[     \frac{\Gamma (\Upsilon/2 - 1/2)}{\Gamma (\Upsilon/2)}\right]^2.
\end{eqnarray}
If $\Upsilon=2$, we recover the SIS model. Thus, by combining Eq.(s) (3) and (19), we may obtain:

\begin{equation}
D_0  = \frac{e^4 \theta_E}{\alpha_{0}^{2} \hbar^2 4\pi \sigma_{ap}^{2}} \Bigg(\frac{\theta_{ap}}{\theta_E}   \Bigg)^{2-\Upsilon}f^{-1}(\Upsilon).
\end{equation}

{In this paper, the factor $\Upsilon$ and the parameter $\gamma$ are approached as free parameters. In addition, the full SGL sample (158 data points) is culled to $N_i$ after the following cuts: $D_0 \pm \sigma_{D_0} > 1$ (non-physical region); the system J0850-0347\footnote{It deviates by more than $5\sigma$ from all the considered models \cite{2018MNRAS.478.5104L}.}; and systems that do not have the corresponding pair of $D_{L_i}$. {Thus, we finish with the following samples: SGL + SNe Ia (total of 89 data points); SGL + ET mock data 1 (total of 97 data points); SGL + ET mock data 2 (total of 69 data points); {\  SGL + LISA (total of 2 points)}. Figure 1 shows the luminosity distances for these samples.}}

\begin{table}
\centering
	\begin{tabular}{|c|c|c|c|} 
	\hline
Data-Set  & $N$ & $\gamma$ & $\Upsilon$ \\ \hline 
SGL + Pantheon & 89 & $-0.04_{-0.04}^{+0.04}$ & $1.96_{-0.06}^{+0.06}$  \\ 
SGL + ET mock data 1 + LISA & 97  & $-0.03_{-0.02}^{+0.02}$ & $2.01_{-0.05}^{+0.05}$  \\
SGL + ET mock data 2 + LISA & 69 & $-0.03_{-0.03}^{+0.03}$ & $1.94_{-0.05}^{+0.06}$  \\ 
\hline    
	\end{tabular}
	\caption{Constraints at $68\%$ CL on the free parameters $\gamma$ and $\Upsilon$.}
\end{table}

\section{Analysis and Discussions}
\label{results}
We used Markov Chain Monte Carlo (MCMC) methods to estimate the posterior probability distribution functions (PDF) of free parameters supported by \texttt{emcee} MCMC sampler \cite{Foreman}. The likelihood is built as follows

\begin{equation}
\mathcal{L} (Data|\Vec{\Theta}) = \prod \frac{1}{\sqrt{2\pi} \sigma_{\mu}} exp \Bigg( -\frac{1}{2} \chi^2   \Bigg),
\end{equation}
where

\begin{equation}
\chi^2 = \sum_{i,j} \left[ (D_{0,i} - \zeta_i)C^{-1}_{T,ij}(D_{0,j} - \zeta_j) \right],
\end{equation}

\begin{equation}
\label{theory_lik}
\zeta_i  \equiv \phi^2 (z_{s,i}) \left[ 1- \frac{(1+z_{s,i})D_{L_{l,i}}}{(1+z_{l,i})D_{L_{s,i}}}\frac{\phi^{1/2}(z_{s,i})}{\phi^{1/2}(z_{l,i})} \right],
\end{equation}
and

\begin{equation}
C_{T,i} = C_{D_{0,i}}+ C_{\zeta,i}.
\end{equation}
$C_{D_{0,i}}$ is the diagonal matrix formed by the squared uncertainties of $D_0$. To obtain $C_{\zeta,i}$ we perform a MonteCarlo sampling of the variable $\zeta_i $ considering a joint multivariate gaussian distribution with mean vector $\bar D_L$ (Eq. \ref{DLave}) and with covariance $cov(\bar D_L, \bar  D_L)$ given by the Eq. (\ref{covDL}). {  The uncertainties of $D_0$ are given by:}

\begin{figure}
    \centering
    \includegraphics[scale=1.0]{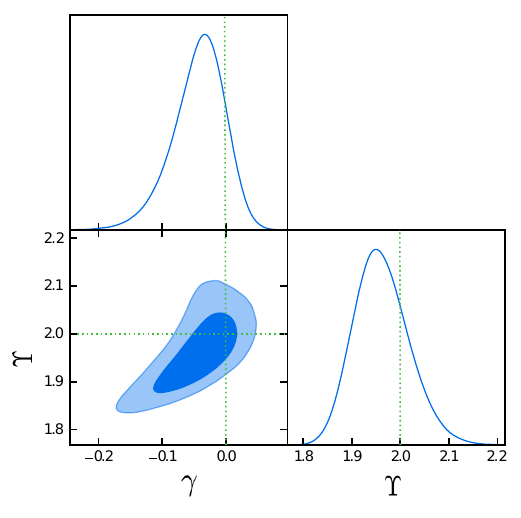}
    \caption{Two-dimensional joint posterior distributions in the $\gamma$-$\Upsilon$ plane from Pantheon + SGL, with the corresponding $68\%$ and $95\%$ CL contours. The green vertical and horizontal dashed lines correspond to the limit $\gamma = 0$ (GR prediction) and $\Upsilon = 2$.}
    \label{sne}
\end{figure}

\begin{figure*}
    \centering
    \includegraphics[scale=1.0]{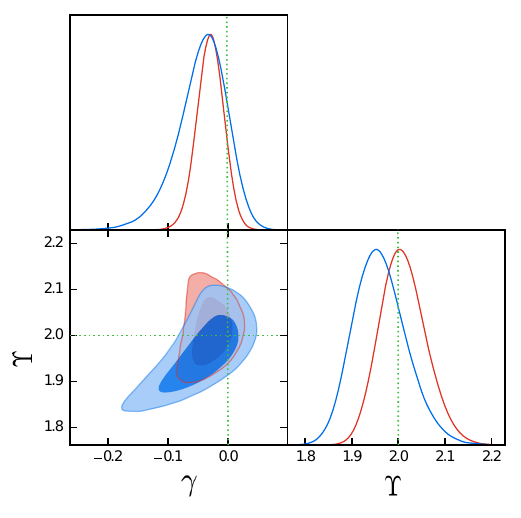}
     \includegraphics[scale=1.0]{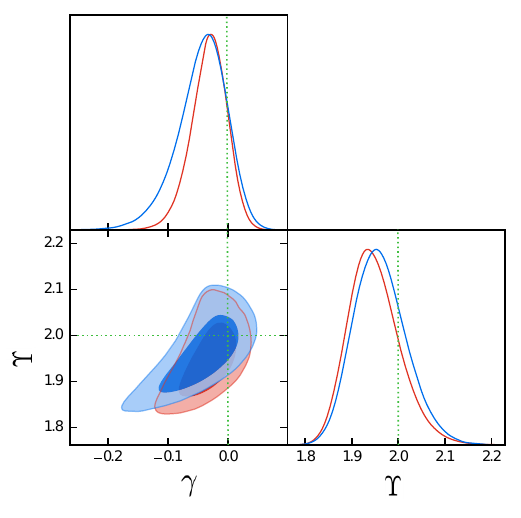}
    \caption{Left panel: Two-dimensional joint posterior distributions in the $\gamma$-$\Upsilon$ plane, with the corresponding $68\%$ and $95\%$ CL contours, obtained from Pantheon + SGL (in blue) and SGL + GWs joint analysis (in red). Here the GWs joint analysis means ET - mock data 1 + LISA.  The green vertical and horizontal dashed lines correspond to the limit $\gamma = 0$ (GR prediction) and $\Upsilon = 2$. Right panel: Same as in left panel, but using ET - mock data 2 + LISA.}
    \label{ET1}
\end{figure*}

\begin{equation}
    \sigma_{D_0}^2 = D_0^2 \left[4 \Bigg( \frac{\sigma_{\sigma_{ap}}}{\sigma_{ap}} \Bigg)^2 + (1-\Upsilon)^2\Bigg(  \frac{\sigma_{\theta_E}}{\theta_E} \Bigg)^2\right].
\end{equation}
 As mentioned before, $\phi (z_s)= 1-\gamma \ln{(1+z_s)}$ and $\phi(z_l) = 1-\gamma \ln{(1+z_l)}$. The PDF posteriori is proportional to the product between the likelihood and the prior, that is,

\begin{equation}
P(\Vec{\Theta} |Data) \propto \mathcal{L} (Data|\Vec{\Theta})\times P_0(\Vec{\Theta}),
\end{equation}
where $\Vec{\Theta} = (\gamma,\Upsilon)$. Nevertheless, we assume flat priors: $-1 \leq \gamma \leq 1$ and $1.5 \leq \Upsilon \leq 2.5$. Following the approach presented by \cite{Grillo}, Einstein’s radius uncertainties follow $\sigma_E = 0.05\theta_E$ ($5\%$ for all systems\footnote{  Even though individual uncertainties are different depending on the survey and on whether the image was taken from the Earth or from space, there is a consensus that in average the relative uncertainty of the Einstein radius is at the level of $5\%$ \cite{2008A&A...477..397G,cao2015}.}). {  In addition, we also follow the procedure of replacing $\sigma_{ap}$ by $\sigma_0$ \cite{cao2015,leo2,leo1,Jorgensen} within the power-law model. The use of $\sigma_0$ makes $D_{0}$ more homogeneous for the sample of lenses located at different redshifts, this occurs because $\sigma_0$ does depend on the measured effective radius ($\theta_E$).}

In this work, we consider two different data combinations, namely, Pantheon + SGL and GWs + SGL, applied to the cosmological model under consideration. {  We emphasize that 89 lensed events are associated with SNe Ia sample, and 97 (69) lensed events with the ET mock data 1 (2), respectively. LISA have only 2 lensed events associated during the MCMC fit.} To derive the constraints on the parameters baseline of this work, we ensure a Gelman - Rubin convergence criterion of $R - 1 < 10^{-3}$~\cite{Gelman} on our chains. 

{   It is worth to point that the method is fully independent of the $M_B$ quantity in supernova data (or on the $M_B-H_0$ tension) \cite{leo1}. In other words,  one does not need to assume a value for the supernova absolute magnitude in Eq. (13) to turn the distance modulus into measurements of the luminosity distance by Eq.(12). This occurs due to the ratio between the luminosity distances presented in Eq.(13). {  We also highlight that the statistical analysis done are independent of $H_0$ and $\Omega_m$, once is not necessary that these parameters pass through the MCMC inference performed from eq. (\ref{theory_lik}).}}

To perform the SS forecasts analysis we need to assume a fiducial cosmology. Thus, the most natural choice is the Planck-$\Lambda$CDM baseline. {  Also note that GWs data modeling, including the instrument noise modeling, everything is considered in generation of the $d_L(z)$ estimates, as described and referenced in section IV B. Thus, the likelihood is built as a direct product of the $d_L(z)$ mock data only, where several systematical and physical effects have already been incorporated in the creation of luminosity distance catalogs used in our analysis following the default specifications and methodologies for the LISA and ET instrument (see Section IV B).} Table 1 summarizes the main results of our statistical analyses.

Fig. \ref{sne} shows the 2D joint posterior distributions at $68\%$~CL and $95\%$~CL in the $\gamma$-$\Upsilon$ plane for Pantheon + SGL. We note $\gamma = -0.04_{-0.04}^{+0.04}$ and $\Upsilon = 1.96_{-0.05}^{+0.06}$ at $68\%$~CL. In Ref. \cite{leo1} was obtained $\gamma = -0.03_{-0.04}^{+0.03}$ at $68\%$~CL. Therefore, both results are in full agreement with each other and with the standard cosmology prediction.

\begin{figure}
    \centering
    \includegraphics[scale=1.0]{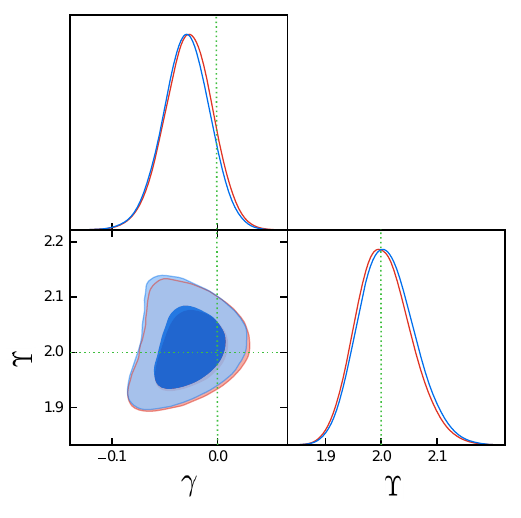}
    \caption{ Two-dimensional joint posterior distributions in the $\gamma$-$\Upsilon$ plane, with the corresponding $68\%$ and $95\%$ CL contours, obtained from SGL + ET mock data 1 + LISA (in blue) and SGL + ET mock data 1 (in red). The green vertical and horizontal dashed lines correspond to the limit $\gamma = 0$ (GR prediction) and $\Upsilon = 2$.}
    \label{LISA_test}
\end{figure}

 From the perspective of forecasts analysis, i.e, SGL + GWs, GWs from ET mock data 1 + LISA combination (see Fig. \ref{ET1} on the left panel, red regions), we find $\gamma = -0.03_{-0.02}^{+0.02}$ and $\Upsilon = 2.01_{-0.05}^{+0.05}$ at $68\%$~CL. This confirms that future distance luminosity measurements from GWs observation can improve the current constraint at $\gamma$ (red regions). In addition, the right panel in Fig. \ref{ET1} shows the 2D joint posterior distributions at $68\%$ CL and $95\%$ CL in the $\gamma$-$\Upsilon$ plane, but from the perspective of the ET mock data 2 + LISA (red regions). In this case, we find  $\gamma = -0.03_{-0.03}^{+0.03}$ and $\Upsilon = 1.94_{-0.05}^{+0.06}$ at $68\%$ CL. The results are very similar to the previous case, both in full agreement with $\Lambda$CDM prediction.

{  Furthermore, we check the effect of SGL + SS  for the LISA and ET separately. Figure \ref{LISA_test} shows the parametric space at $68\%$~CL and $95\%$~CL in the $\gamma$-$\Upsilon$ plane from SGL + ET mock data 1 + LISA (in blue) and SGL + ET mock data 1 (in red). Thus, we can quantify how much LISA can influence in the observational constrains. In LISA sample, we have only two pairs lensed events located at $z = 0.497 - 2.092$ and $0.491 - 1.232$. Once our methodology only takes into account the geometrical distance measurements of the SS event, we can see that the joint analysis LISA+SGL system sample does not contribute much for a better accuracy of the parameter space of the model. Thus, our analysis with and without LISA sample represents the same impact. Because the SS event detection from LISA perspective are very low than ET, a more effective way to perform cosmological tests will be use SGL of gravitational waves through a direct and individual analysis of the merging of massive black hole binaries in the context of the LISA mission \cite{Sereno_2010}, of which in principle can done to constraints cosmological parameters with and without the identification of their electromagnetic counterparts \cite{Sereno_2011}. Several works have demonstrated the potential of these systems to infer cosmological constraints (see \cite{huang2023measuring, Liu_2019,Hou_2020,lin2023strong, Ali_2023} for a few example). Therefore, in that regard, our eq. (26) will have to be adapted to take into account the physical-astrophysical information of the statistics of gravitational lenses of each system instead of evaluating only the simple geometrical distance measurements. These perspectives are beyond the discussion and presentation of our current methodology. We leave these details for a future work.}
\\

\section{Conclusions}
\label{final}

The space-time variations in fundamental constants are expected to naturally arise from the interactions of new low-mass particles appearing, for example, in theories of dark energy and modified gravity. In this work, we  investigated the  potentialities of future gravitational-wave standard sirens for constraining  a possible  time variation of the fine structure constant within the runaway dilaton model. The forecasts analysis was based on the generation of some standard sirens catalogs within the perspective of two future gravitational waves observatories, namely, Einstein Telescope and LISA,  jointly with current strong gravitational lensing systems. Our results showed no significant evidence for a varying-$\alpha$ model, being the specific framework tested in this work in full agreement with $\Lambda$-cosmology.

Therefore, we  obtained that future measurements of the luminosity distances of GWs sources plus SGL systems will be at least competitive with the current analyses by using SNe Ia and SGL data (see Figs. 2 and 3). 
{  On the other hand, some caveats are important to mention. The new cosmological test proposed here may contain some sources of systematic effects, of which is not the main purpose of this present work to check in detail, but are assumptions assumed. For instance, the power-law lens model and the SS events distribution rate. Because it is a forecast analysis in relation to the SS sample, and not a real data analysis, we do not expect main changes in the accuracy inferred in the main parameters, but they are certainly important systematic points to be addressed with future events based on real data.}

In the near future, our approach will be able to put tighter limits on $\Delta \alpha /\alpha$ by using GW data and thousands of strong lensing systems that will be discovered by the optical and infrared data from the EUCLID mission, Vera Rubin LSST, and Nancy Grace Roman space telescope. A natural extension of this work will be to consider the role of cosmic curvature in our current results. Moreover,  similar analyses could also be done with BAO measurements. Such analysis is left as future considerations to be developed.
\\

\section{Acknowledgments}
\textbf{The authors thank the referees for useful comments and sugestions.} RCN thanks the financial support from the Funda\c{c}\~{a}o de Amparo \`{a} Pesquisa do Estado de S\~{a}o Paulo (FAPESP, S\~{a}o Paulo Research Foundation) under the project No. 2018/18036-5, Conselho Nacional de Desenvolvimento Cient\'{i}fico e Tecnologico (CNPq, National Council for Scientific and Technological Development) for partial financial support under the project No. 304306/2022-3, and the Fundação de Amparo à pesquisa do Estado do RS (FAPERGS, Research Support Foundation of the State of RS) for partial financial support under the project No. 23/2551-0000848-3. RFL thanks to CNPq.
\\

\section{Data Availability Statement}
Data will be made available on reasonable request.

\bibliography{bibly}

\begin{thebibliography}{147}
\expandafter\ifx\csname natexlab\endcsname\relax\def\natexlab#1{#1}\fi
\expandafter\ifx\csname bibnamefont\endcsname\relax
  \def\bibnamefont#1{#1}\fi
\expandafter\ifx\csname bibfnamefont\endcsname\relax
  \def\bibfnamefont#1{#1}\fi
\expandafter\ifx\csname citenamefont\endcsname\relax
  \def\citenamefont#1{#1}\fi
\expandafter\ifx\csname url\endcsname\relax
  \def\url#1{\texttt{#1}}\fi
\expandafter\ifx\csname urlprefix\endcsname\relax\def\urlprefix{URL }\fi
\providecommand{\bibinfo}[2]{#2}
\providecommand{\eprint}[2][]{\url{#2}}

\bibitem[{\citenamefont{Dirac}(1937)}]{Dirac1}
\bibinfo{author}{\bibfnamefont{P.}~\bibnamefont{Dirac}},
  \bibinfo{journal}{Nature} \textbf{\bibinfo{volume}{139}}
  (\bibinfo{year}{1937}),
  \urlprefix\url{https://www.nature.com/articles/139323a0#citeas}.

\bibitem[{\citenamefont{{Milne}}(1935)}]{Milne1935}
\bibinfo{author}{\bibfnamefont{E.~A.} \bibnamefont{{Milne}}},
  \emph{\bibinfo{title}{{Relativity, gravitation and world-structure}}}
  (\bibinfo{year}{1935}).

\bibitem[{\citenamefont{{Jordan}}(1937)}]{jordan1937}
\bibinfo{author}{\bibfnamefont{P.}~\bibnamefont{{Jordan}}},
  \bibinfo{journal}{Naturwissenschaften} \textbf{\bibinfo{volume}{25}},
  \bibinfo{pages}{513} (\bibinfo{year}{1937}).

\bibitem[{\citenamefont{{Jordan}}(1939)}]{jordan1939}
\bibinfo{author}{\bibfnamefont{P.}~\bibnamefont{{Jordan}}},
  \bibinfo{journal}{Zeitschrift fur Physik} \textbf{\bibinfo{volume}{113}},
  \bibinfo{pages}{660} (\bibinfo{year}{1939}).

\bibitem[{\citenamefont{{Burrage} and
  {Dombrowski}}(2020)}]{2020JCAP...07..060B}
\bibinfo{author}{\bibfnamefont{C.}~\bibnamefont{{Burrage}}} \bibnamefont{and}
  \bibinfo{author}{\bibfnamefont{J.}~\bibnamefont{{Dombrowski}}},
  \bibinfo{journal}{\jcap} \textbf{\bibinfo{volume}{2020}}, \bibinfo{eid}{060}
  (\bibinfo{year}{2020}), \eprint{2004.14260}.

\bibitem[{\citenamefont{{Clifton} et~al.}(2012)\citenamefont{{Clifton},
  {Ferreira}, {Padilla}, and {Skordis}}}]{2012PhR...513....1C}
\bibinfo{author}{\bibfnamefont{T.}~\bibnamefont{{Clifton}}},
  \bibinfo{author}{\bibfnamefont{P.~G.} \bibnamefont{{Ferreira}}},
  \bibinfo{author}{\bibfnamefont{A.}~\bibnamefont{{Padilla}}},
  \bibnamefont{and}
  \bibinfo{author}{\bibfnamefont{C.}~\bibnamefont{{Skordis}}},
  \bibinfo{journal}{\physrep} \textbf{\bibinfo{volume}{513}},
  \bibinfo{pages}{1} (\bibinfo{year}{2012}), \eprint{1106.2476}.

\bibitem[{\citenamefont{Brans and Dicke}(1961)}]{PhysRev.124.925}
\bibinfo{author}{\bibfnamefont{C.}~\bibnamefont{Brans}} \bibnamefont{and}
  \bibinfo{author}{\bibfnamefont{R.~H.} \bibnamefont{Dicke}},
  \bibinfo{journal}{Phys. Rev.} \textbf{\bibinfo{volume}{124}},
  \bibinfo{pages}{925} (\bibinfo{year}{1961}),
  \urlprefix\url{https://link.aps.org/doi/10.1103/PhysRev.124.925}.

\bibitem[{\citenamefont{Gamow}(1967)}]{PhysRevLett.19.759}
\bibinfo{author}{\bibfnamefont{G.}~\bibnamefont{Gamow}},
  \bibinfo{journal}{Phys. Rev. Lett.} \textbf{\bibinfo{volume}{19}},
  \bibinfo{pages}{759} (\bibinfo{year}{1967}),
  \urlprefix\url{https://link.aps.org/doi/10.1103/PhysRevLett.19.759}.

\bibitem[{\citenamefont{{Albrecht} and {Magueijo}}(1999)}]{1999PhRvD..59d3516A}
\bibinfo{author}{\bibfnamefont{A.}~\bibnamefont{{Albrecht}}} \bibnamefont{and}
  \bibinfo{author}{\bibfnamefont{J.}~\bibnamefont{{Magueijo}}},
  \bibinfo{journal}{\prd} \textbf{\bibinfo{volume}{59}}, \bibinfo{eid}{043516}
  (\bibinfo{year}{1999}), \eprint{astro-ph/9811018}.

\bibitem[{\citenamefont{{Ellis} and {Uzan}}(2005)}]{2005AmJPh..73..240E}
\bibinfo{author}{\bibfnamefont{G.~F.~R.} \bibnamefont{{Ellis}}}
  \bibnamefont{and} \bibinfo{author}{\bibfnamefont{J.-P.}
  \bibnamefont{{Uzan}}}, \bibinfo{journal}{American Journal of Physics}
  \textbf{\bibinfo{volume}{73}}, \bibinfo{pages}{240} (\bibinfo{year}{2005}),
  \eprint{gr-qc/0305099}.

\bibitem[{\citenamefont{{Cao} et~al.}(2018)\citenamefont{{Cao}, {Qi},
  {Biesiada}, {Zheng}, {Xu}, and {Zhu}}}]{2018ApJ...867...50C}
\bibinfo{author}{\bibfnamefont{S.}~\bibnamefont{{Cao}}},
  \bibinfo{author}{\bibfnamefont{J.}~\bibnamefont{{Qi}}},
  \bibinfo{author}{\bibfnamefont{M.}~\bibnamefont{{Biesiada}}},
  \bibinfo{author}{\bibfnamefont{X.}~\bibnamefont{{Zheng}}},
  \bibinfo{author}{\bibfnamefont{T.}~\bibnamefont{{Xu}}}, \bibnamefont{and}
  \bibinfo{author}{\bibfnamefont{Z.-H.} \bibnamefont{{Zhu}}},
  \bibinfo{journal}{\apj} \textbf{\bibinfo{volume}{867}}, \bibinfo{eid}{50}
  (\bibinfo{year}{2018}), \eprint{1810.01287}.

\bibitem[{\citenamefont{{Cao} et~al.}(2017)\citenamefont{{Cao}, {Biesiada},
  {Jackson}, {Zheng}, {Zhao}, and {Zhu}}}]{2017JCAP...02..012C}
\bibinfo{author}{\bibfnamefont{S.}~\bibnamefont{{Cao}}},
  \bibinfo{author}{\bibfnamefont{M.}~\bibnamefont{{Biesiada}}},
  \bibinfo{author}{\bibfnamefont{J.}~\bibnamefont{{Jackson}}},
  \bibinfo{author}{\bibfnamefont{X.}~\bibnamefont{{Zheng}}},
  \bibinfo{author}{\bibfnamefont{Y.}~\bibnamefont{{Zhao}}}, \bibnamefont{and}
  \bibinfo{author}{\bibfnamefont{Z.-H.} \bibnamefont{{Zhu}}},
  \bibinfo{journal}{\jcap} \textbf{\bibinfo{volume}{2017}}, \bibinfo{eid}{012}
  (\bibinfo{year}{2017}), \eprint{1609.08748}.

\bibitem[{\citenamefont{{Mendon{\c{c}}a}
  et~al.}(2021)\citenamefont{{Mendon{\c{c}}a}, {Bora}, {Holanda}, {Desai}, and
  {Pereira}}}]{2021JCAP...11..034M}
\bibinfo{author}{\bibfnamefont{I.~E.~C.~R.} \bibnamefont{{Mendon{\c{c}}a}}},
  \bibinfo{author}{\bibfnamefont{K.}~\bibnamefont{{Bora}}},
  \bibinfo{author}{\bibfnamefont{R.~F.~L.} \bibnamefont{{Holanda}}},
  \bibinfo{author}{\bibfnamefont{S.}~\bibnamefont{{Desai}}}, \bibnamefont{and}
  \bibinfo{author}{\bibfnamefont{S.~H.} \bibnamefont{{Pereira}}},
  \bibinfo{journal}{\jcap} \textbf{\bibinfo{volume}{2021}}, \bibinfo{eid}{034}
  (\bibinfo{year}{2021}), \eprint{2109.14512}.

\bibitem[{\citenamefont{{Liu} et~al.}(2021{\natexlab{a}})\citenamefont{{Liu},
  {Cao}, {Biesiada}, {Liu}, {Lian}, and {Zhang}}}]{2021MNRAS.506.2181L}
\bibinfo{author}{\bibfnamefont{T.}~\bibnamefont{{Liu}}},
  \bibinfo{author}{\bibfnamefont{S.}~\bibnamefont{{Cao}}},
  \bibinfo{author}{\bibfnamefont{M.}~\bibnamefont{{Biesiada}}},
  \bibinfo{author}{\bibfnamefont{Y.}~\bibnamefont{{Liu}}},
  \bibinfo{author}{\bibfnamefont{Y.}~\bibnamefont{{Lian}}}, \bibnamefont{and}
  \bibinfo{author}{\bibfnamefont{Y.}~\bibnamefont{{Zhang}}},
  \bibinfo{journal}{\mnras} \textbf{\bibinfo{volume}{506}},
  \bibinfo{pages}{2181} (\bibinfo{year}{2021}{\natexlab{a}}),
  \eprint{2106.15145}.

\bibitem[{\citenamefont{Uzan}(2011)}]{Uzan2011}
\bibinfo{author}{\bibfnamefont{J.-P.} \bibnamefont{Uzan}},
  \bibinfo{journal}{Living Reviews in Relativity} \textbf{\bibinfo{volume}{14}}
  (\bibinfo{year}{2011}), ISSN \bibinfo{issn}{1433-8351},
  \urlprefix\url{http://dx.doi.org/10.12942/lrr-2011-2}.

\bibitem[{\citenamefont{Martins}(2017)}]{Martins2017}
\bibinfo{author}{\bibfnamefont{C.~J. A.~P.} \bibnamefont{Martins}},
  \bibinfo{journal}{Reports on Progress in Physics}
  \textbf{\bibinfo{volume}{80}}, \bibinfo{pages}{126902}
  (\bibinfo{year}{2017}), ISSN \bibinfo{issn}{1361-6633},
  \urlprefix\url{http://dx.doi.org/10.1088/1361-6633/aa860e}.

\bibitem[{\citenamefont{Damour}(2003)}]{Damour:2002vu}
\bibinfo{author}{\bibfnamefont{T.}~\bibnamefont{Damour}},
  \bibinfo{journal}{Astrophys. Space Sci.} \textbf{\bibinfo{volume}{283}},
  \bibinfo{pages}{445} (\bibinfo{year}{2003}), \eprint{gr-qc/0210059}.

\bibitem[{\citenamefont{et~al}(2008)}]{Rosenband}
\bibinfo{author}{\bibfnamefont{R.~T.} \bibnamefont{et~al}},
  \bibinfo{journal}{Science} \textbf{\bibinfo{volume}{319}},
  \bibinfo{pages}{1808} (\bibinfo{year}{2008}).

\bibitem[{\citenamefont{{Fujii} and {Maeda}}(2003)}]{2003sttg.book.....F}
\bibinfo{author}{\bibfnamefont{Y.}~\bibnamefont{{Fujii}}} \bibnamefont{and}
  \bibinfo{author}{\bibfnamefont{K.-I.} \bibnamefont{{Maeda}}},
  \emph{\bibinfo{title}{{The Scalar-Tensor Theory of Gravitation}}}
  (\bibinfo{year}{2003}).

\bibitem[{\citenamefont{{Naruko} et~al.}(2016)\citenamefont{{Naruko},
  {Yoshida}, and {Mukohyama}}}]{2016CQGra..33iLT01N}
\bibinfo{author}{\bibfnamefont{A.}~\bibnamefont{{Naruko}}},
  \bibinfo{author}{\bibfnamefont{D.}~\bibnamefont{{Yoshida}}},
  \bibnamefont{and}
  \bibinfo{author}{\bibfnamefont{S.}~\bibnamefont{{Mukohyama}}},
  \bibinfo{journal}{Classical and Quantum Gravity}
  \textbf{\bibinfo{volume}{33}}, \bibinfo{eid}{09LT01} (\bibinfo{year}{2016}),
  \eprint{1512.06977}.

\bibitem[{\citenamefont{{Hees} et~al.}(2014)\citenamefont{{Hees}, {Minazzoli},
  and {Larena}}}]{hees}
\bibinfo{author}{\bibfnamefont{A.}~\bibnamefont{{Hees}}},
  \bibinfo{author}{\bibfnamefont{O.}~\bibnamefont{{Minazzoli}}},
  \bibnamefont{and} \bibinfo{author}{\bibfnamefont{J.}~\bibnamefont{{Larena}}},
  \bibinfo{journal}{\prd} \textbf{\bibinfo{volume}{90}}, \bibinfo{eid}{124064}
  (\bibinfo{year}{2014}), \eprint{1406.6187}.

\bibitem[{\citenamefont{de~Bruck et~al.}(2015)\citenamefont{de~Bruck, Mifsud,
  and Nunes}}]{Bruck2015}
\bibinfo{author}{\bibfnamefont{C.~v.} \bibnamefont{de~Bruck}},
  \bibinfo{author}{\bibfnamefont{J.}~\bibnamefont{Mifsud}}, \bibnamefont{and}
  \bibinfo{author}{\bibfnamefont{N.~J.} \bibnamefont{Nunes}},
  \bibinfo{journal}{Journal of Cosmology and Astroparticle Physics}
  \textbf{\bibinfo{volume}{2015}}, \bibinfo{pages}{018–018}
  (\bibinfo{year}{2015}), ISSN \bibinfo{issn}{1475-7516},
  \urlprefix\url{http://dx.doi.org/10.1088/1475-7516/2015/12/018}.

\bibitem[{\citenamefont{Nunes et~al.}(2017)\citenamefont{Nunes, Bonilla, Pan,
  and Saridakis}}]{Nunes2017}
\bibinfo{author}{\bibfnamefont{R.~C.} \bibnamefont{Nunes}},
  \bibinfo{author}{\bibfnamefont{A.}~\bibnamefont{Bonilla}},
  \bibinfo{author}{\bibfnamefont{S.}~\bibnamefont{Pan}}, \bibnamefont{and}
  \bibinfo{author}{\bibfnamefont{E.~N.} \bibnamefont{Saridakis}},
  \bibinfo{journal}{The European Physical Journal C}
  \textbf{\bibinfo{volume}{77}} (\bibinfo{year}{2017}), ISSN
  \bibinfo{issn}{1434-6052},
  \urlprefix\url{http://dx.doi.org/10.1140/epjc/s10052-017-4798-5}.

\bibitem[{\citenamefont{Said et~al.}(2020)\citenamefont{Said, Mifsud,
  Parkinson, Saridakis, Sultana, and Adami}}]{Said2020}
\bibinfo{author}{\bibfnamefont{J.~L.} \bibnamefont{Said}},
  \bibinfo{author}{\bibfnamefont{J.}~\bibnamefont{Mifsud}},
  \bibinfo{author}{\bibfnamefont{D.}~\bibnamefont{Parkinson}},
  \bibinfo{author}{\bibfnamefont{E.~N.} \bibnamefont{Saridakis}},
  \bibinfo{author}{\bibfnamefont{J.}~\bibnamefont{Sultana}}, \bibnamefont{and}
  \bibinfo{author}{\bibfnamefont{K.~Z.} \bibnamefont{Adami}},
  \bibinfo{journal}{Journal of Cosmology and Astroparticle Physics}
  \textbf{\bibinfo{volume}{2020}}, \bibinfo{pages}{047–047}
  (\bibinfo{year}{2020}), ISSN \bibinfo{issn}{1475-7516},
  \urlprefix\url{http://dx.doi.org/10.1088/1475-7516/2020/11/047}.

\bibitem[{\citenamefont{Fritzsch et~al.}(2017)\citenamefont{Fritzsch, Solà,
  and Nunes}}]{Nunes2017b}
\bibinfo{author}{\bibfnamefont{H.}~\bibnamefont{Fritzsch}},
  \bibinfo{author}{\bibfnamefont{J.}~\bibnamefont{Solà}}, \bibnamefont{and}
  \bibinfo{author}{\bibfnamefont{R.~C.} \bibnamefont{Nunes}},
  \bibinfo{journal}{The European Physical Journal C}
  \textbf{\bibinfo{volume}{77}} (\bibinfo{year}{2017}), ISSN
  \bibinfo{issn}{1434-6052},
  \urlprefix\url{http://dx.doi.org/10.1140/epjc/s10052-017-4714-z}.

\bibitem[{\citenamefont{{Bekenstein}}(1982)}]{1982PhRvD..25.1527B}
\bibinfo{author}{\bibfnamefont{J.~D.} \bibnamefont{{Bekenstein}}},
  \bibinfo{journal}{\prd} \textbf{\bibinfo{volume}{25}}, \bibinfo{pages}{1527}
  (\bibinfo{year}{1982}).

\bibitem[{\citenamefont{{Barrow} and {Lip}}(2012)}]{2012PhRvD..85b3514B}
\bibinfo{author}{\bibfnamefont{J.~D.} \bibnamefont{{Barrow}}} \bibnamefont{and}
  \bibinfo{author}{\bibfnamefont{S.~Z.~W.} \bibnamefont{{Lip}}},
  \bibinfo{journal}{\prd} \textbf{\bibinfo{volume}{85}}, \bibinfo{eid}{023514}
  (\bibinfo{year}{2012}), \eprint{1110.3120}.

\bibitem[{\citenamefont{{Overduin} and {Wesson}}(1997)}]{1997PhR...283..303O}
\bibinfo{author}{\bibfnamefont{J.~M.} \bibnamefont{{Overduin}}}
  \bibnamefont{and} \bibinfo{author}{\bibfnamefont{P.~S.}
  \bibnamefont{{Wesson}}}, \bibinfo{journal}{\physrep}
  \textbf{\bibinfo{volume}{283}}, \bibinfo{pages}{303} (\bibinfo{year}{1997}),
  \eprint{gr-qc/9805018}.

\bibitem[{\citenamefont{{Martins} et~al.}(2016)\citenamefont{{Martins},
  {Pinho}, {Carreira}, {Gusart}, {L{\'o}pez}, and
  {Rocha}}}]{2016PhRvD..93b3506M}
\bibinfo{author}{\bibfnamefont{C.~J.~A.~P.} \bibnamefont{{Martins}}},
  \bibinfo{author}{\bibfnamefont{A.~M.~M.} \bibnamefont{{Pinho}}},
  \bibinfo{author}{\bibfnamefont{P.}~\bibnamefont{{Carreira}}},
  \bibinfo{author}{\bibfnamefont{A.}~\bibnamefont{{Gusart}}},
  \bibinfo{author}{\bibfnamefont{J.}~\bibnamefont{{L{\'o}pez}}},
  \bibnamefont{and} \bibinfo{author}{\bibfnamefont{C.~I.~S.~A.}
  \bibnamefont{{Rocha}}}, \bibinfo{journal}{\prd}
  \textbf{\bibinfo{volume}{93}}, \bibinfo{eid}{023506} (\bibinfo{year}{2016}),
  \eprint{1601.02950}.

\bibitem[{\citenamefont{{Martins} and {Pinho}}(2015)}]{2015PhRvD..91j3501M}
\bibinfo{author}{\bibfnamefont{C.~J.~A.~P.} \bibnamefont{{Martins}}}
  \bibnamefont{and} \bibinfo{author}{\bibfnamefont{A.~M.~M.}
  \bibnamefont{{Pinho}}}, \bibinfo{journal}{\prd}
  \textbf{\bibinfo{volume}{91}}, \bibinfo{eid}{103501} (\bibinfo{year}{2015}),
  \eprint{1505.02196}.

\bibitem[{\citenamefont{{Mosquera} and
  {Civitarese}}(2017)}]{2017PhRvC..96d5802M}
\bibinfo{author}{\bibfnamefont{M.~E.} \bibnamefont{{Mosquera}}}
  \bibnamefont{and}
  \bibinfo{author}{\bibfnamefont{O.}~\bibnamefont{{Civitarese}}},
  \bibinfo{journal}{\prc} \textbf{\bibinfo{volume}{96}}, \bibinfo{eid}{045802}
  (\bibinfo{year}{2017}).

\bibitem[{\citenamefont{{Liu} et~al.}(2021{\natexlab{b}})\citenamefont{{Liu},
  {Liu}, {Zhang}, {Zhai}, and {Bora}}}]{2021ApJ...922...19L}
\bibinfo{author}{\bibfnamefont{Z.-E.} \bibnamefont{{Liu}}},
  \bibinfo{author}{\bibfnamefont{W.-F.} \bibnamefont{{Liu}}},
  \bibinfo{author}{\bibfnamefont{T.-J.} \bibnamefont{{Zhang}}},
  \bibinfo{author}{\bibfnamefont{Z.-X.} \bibnamefont{{Zhai}}},
  \bibnamefont{and} \bibinfo{author}{\bibfnamefont{K.}~\bibnamefont{{Bora}}},
  \bibinfo{journal}{\apj} \textbf{\bibinfo{volume}{922}}, \bibinfo{eid}{19}
  (\bibinfo{year}{2021}{\natexlab{b}}), \eprint{2109.00134}.

\bibitem[{\citenamefont{{Landau}}(2020)}]{Landau:2020vkr}
\bibinfo{author}{\bibfnamefont{S.~J.} \bibnamefont{{Landau}}},
  \bibinfo{journal}{IAU Symposium} \textbf{\bibinfo{volume}{357}},
  \bibinfo{pages}{45} (\bibinfo{year}{2020}), \eprint{2002.00095}.

\bibitem[{\citenamefont{{Bainbridge} et~al.}(2017)\citenamefont{{Bainbridge},
  {Barstow}, {Reindl}, {Tchang-Brillet}, {Ayres}, {Webb}, {Barrow}, {Hu},
  {Holberg}, {Preval} et~al.}}]{Bainbridge:2017lsj}
\bibinfo{author}{\bibfnamefont{M.}~\bibnamefont{{Bainbridge}}},
  \bibinfo{author}{\bibfnamefont{M.}~\bibnamefont{{Barstow}}},
  \bibinfo{author}{\bibfnamefont{N.}~\bibnamefont{{Reindl}}},
  \bibinfo{author}{\bibfnamefont{W.~{\"U}.} \bibnamefont{{Tchang-Brillet}}},
  \bibinfo{author}{\bibfnamefont{T.}~\bibnamefont{{Ayres}}},
  \bibinfo{author}{\bibfnamefont{J.}~\bibnamefont{{Webb}}},
  \bibinfo{author}{\bibfnamefont{J.}~\bibnamefont{{Barrow}}},
  \bibinfo{author}{\bibfnamefont{J.}~\bibnamefont{{Hu}}},
  \bibinfo{author}{\bibfnamefont{J.}~\bibnamefont{{Holberg}}},
  \bibinfo{author}{\bibfnamefont{S.}~\bibnamefont{{Preval}}},
  \bibnamefont{et~al.}, \bibinfo{journal}{Universe}
  \textbf{\bibinfo{volume}{3}}, \bibinfo{pages}{32} (\bibinfo{year}{2017}),
  \eprint{1702.01757}.

\bibitem[{\citenamefont{{Wilczynska} et~al.}(2020)\citenamefont{{Wilczynska},
  {Webb}, {Bainbridge}, {Barrow}, {Bosman}, {Carswell}, {Dabrowski}, {Dumont},
  {Lee}, {Leite} et~al.}}]{Wilczynska}
\bibinfo{author}{\bibfnamefont{M.~R.} \bibnamefont{{Wilczynska}}},
  \bibinfo{author}{\bibfnamefont{J.~K.} \bibnamefont{{Webb}}},
  \bibinfo{author}{\bibfnamefont{M.}~\bibnamefont{{Bainbridge}}},
  \bibinfo{author}{\bibfnamefont{J.~D.} \bibnamefont{{Barrow}}},
  \bibinfo{author}{\bibfnamefont{S.~E.~I.} \bibnamefont{{Bosman}}},
  \bibinfo{author}{\bibfnamefont{R.~F.} \bibnamefont{{Carswell}}},
  \bibinfo{author}{\bibfnamefont{M.~P.} \bibnamefont{{Dabrowski}}},
  \bibinfo{author}{\bibfnamefont{V.}~\bibnamefont{{Dumont}}},
  \bibinfo{author}{\bibfnamefont{C.-C.} \bibnamefont{{Lee}}},
  \bibinfo{author}{\bibfnamefont{A.~C.} \bibnamefont{{Leite}}},
  \bibnamefont{et~al.}, \bibinfo{journal}{Science Advances}
  \textbf{\bibinfo{volume}{6}}, \bibinfo{pages}{eaay9672}
  (\bibinfo{year}{2020}), \eprint{2003.07627}.

\bibitem[{\citenamefont{{Webb} et~al.}(1999)\citenamefont{{Webb}, {Flambaum},
  {Churchill}, {Drinkwater}, and {Barrow}}}]{webb1999}
\bibinfo{author}{\bibfnamefont{J.~K.} \bibnamefont{{Webb}}},
  \bibinfo{author}{\bibfnamefont{V.~V.} \bibnamefont{{Flambaum}}},
  \bibinfo{author}{\bibfnamefont{C.~W.} \bibnamefont{{Churchill}}},
  \bibinfo{author}{\bibfnamefont{M.~J.} \bibnamefont{{Drinkwater}}},
  \bibnamefont{and} \bibinfo{author}{\bibfnamefont{J.~D.}
  \bibnamefont{{Barrow}}}, \bibinfo{journal}{\prl}
  \textbf{\bibinfo{volume}{82}}, \bibinfo{pages}{884} (\bibinfo{year}{1999}),
  \eprint{astro-ph/9803165}.

\bibitem[{\citenamefont{{Ubachs}}(2018)}]{Ubachs:2017zmg}
\bibinfo{author}{\bibfnamefont{W.}~\bibnamefont{{Ubachs}}},
  \bibinfo{journal}{\ssr} \textbf{\bibinfo{volume}{214}}, \bibinfo{eid}{3}
  (\bibinfo{year}{2018}), \eprint{1709.07704}.

\bibitem[{\citenamefont{{Milakovi{\'c}}
  et~al.}(2021{\natexlab{a}})\citenamefont{{Milakovi{\'c}}, {Lee}, {Carswell},
  {Webb}, {Molaro}, and {Pasquini}}}]{2021MNRAS.500....1M}
\bibinfo{author}{\bibfnamefont{D.}~\bibnamefont{{Milakovi{\'c}}}},
  \bibinfo{author}{\bibfnamefont{C.-C.} \bibnamefont{{Lee}}},
  \bibinfo{author}{\bibfnamefont{R.~F.} \bibnamefont{{Carswell}}},
  \bibinfo{author}{\bibfnamefont{J.~K.} \bibnamefont{{Webb}}},
  \bibinfo{author}{\bibfnamefont{P.}~\bibnamefont{{Molaro}}}, \bibnamefont{and}
  \bibinfo{author}{\bibfnamefont{L.}~\bibnamefont{{Pasquini}}},
  \bibinfo{journal}{\mnras} \textbf{\bibinfo{volume}{500}}, \bibinfo{pages}{1}
  (\bibinfo{year}{2021}{\natexlab{a}}), \eprint{2008.10619}.

\bibitem[{\citenamefont{{Lee} et~al.}(2021)\citenamefont{{Lee}, {Webb},
  {Milakovi{\'c}}, and {Carswell}}}]{Lee}
\bibinfo{author}{\bibfnamefont{C.-C.} \bibnamefont{{Lee}}},
  \bibinfo{author}{\bibfnamefont{J.~K.} \bibnamefont{{Webb}}},
  \bibinfo{author}{\bibfnamefont{D.}~\bibnamefont{{Milakovi{\'c}}}},
  \bibnamefont{and} \bibinfo{author}{\bibfnamefont{R.~F.}
  \bibnamefont{{Carswell}}}, \bibinfo{journal}{\mnras}
  \textbf{\bibinfo{volume}{507}}, \bibinfo{pages}{27} (\bibinfo{year}{2021}),
  \eprint{2102.11648}.

\bibitem[{\citenamefont{{Galli}}(2013)}]{galli}
\bibinfo{author}{\bibfnamefont{S.}~\bibnamefont{{Galli}}},
  \bibinfo{journal}{\prd} \textbf{\bibinfo{volume}{87}}, \bibinfo{eid}{123516}
  (\bibinfo{year}{2013}), \eprint{1212.1075}.

\bibitem[{\citenamefont{{Clara} and {Martins}}(2020)}]{BBN}
\bibinfo{author}{\bibfnamefont{M.~T.} \bibnamefont{{Clara}}} \bibnamefont{and}
  \bibinfo{author}{\bibfnamefont{C.~J.~A.~P.} \bibnamefont{{Martins}}},
  \bibinfo{journal}{\aap} \textbf{\bibinfo{volume}{633}}, \bibinfo{eid}{L11}
  (\bibinfo{year}{2020}), \eprint{2001.01787}.

\bibitem[{\citenamefont{{Hees} et~al.}(2020)\citenamefont{{Hees}, {Do},
  {Roberts}, {Ghez}, {Nishiyama}, {Bentley}, {Gautam}, {Jia}, {Kara}, {Lu}
  et~al.}}]{Hees:2020gda}
\bibinfo{author}{\bibfnamefont{A.}~\bibnamefont{{Hees}}},
  \bibinfo{author}{\bibfnamefont{T.}~\bibnamefont{{Do}}},
  \bibinfo{author}{\bibfnamefont{B.~M.} \bibnamefont{{Roberts}}},
  \bibinfo{author}{\bibfnamefont{A.~M.} \bibnamefont{{Ghez}}},
  \bibinfo{author}{\bibfnamefont{S.}~\bibnamefont{{Nishiyama}}},
  \bibinfo{author}{\bibfnamefont{R.~O.} \bibnamefont{{Bentley}}},
  \bibinfo{author}{\bibfnamefont{A.~K.} \bibnamefont{{Gautam}}},
  \bibinfo{author}{\bibfnamefont{S.}~\bibnamefont{{Jia}}},
  \bibinfo{author}{\bibfnamefont{T.}~\bibnamefont{{Kara}}},
  \bibinfo{author}{\bibfnamefont{J.~R.} \bibnamefont{{Lu}}},
  \bibnamefont{et~al.}, \bibinfo{journal}{\prl} \textbf{\bibinfo{volume}{124}},
  \bibinfo{eid}{081101} (\bibinfo{year}{2020}), \eprint{2002.11567}.

\bibitem[{\citenamefont{{Milakovi{\'c}}
  et~al.}(2021{\natexlab{b}})\citenamefont{{Milakovi{\'c}}, {Lee}, {Carswell},
  {Webb}, {Molaro}, and {Pasquini}}}]{Milakovic:2020tvq}
\bibinfo{author}{\bibfnamefont{D.}~\bibnamefont{{Milakovi{\'c}}}},
  \bibinfo{author}{\bibfnamefont{C.-C.} \bibnamefont{{Lee}}},
  \bibinfo{author}{\bibfnamefont{R.~F.} \bibnamefont{{Carswell}}},
  \bibinfo{author}{\bibfnamefont{J.~K.} \bibnamefont{{Webb}}},
  \bibinfo{author}{\bibfnamefont{P.}~\bibnamefont{{Molaro}}}, \bibnamefont{and}
  \bibinfo{author}{\bibfnamefont{L.}~\bibnamefont{{Pasquini}}},
  \bibinfo{journal}{\mnras} \textbf{\bibinfo{volume}{500}}, \bibinfo{pages}{1}
  (\bibinfo{year}{2021}{\natexlab{b}}), \eprint{2008.10619}.

\bibitem[{\citenamefont{{Kraiselburd} et~al.}(2018)\citenamefont{{Kraiselburd},
  {Castillo}, {Mosquera}, and {Vucetich}}}]{Kraiselburd:2018uac}
\bibinfo{author}{\bibfnamefont{L.}~\bibnamefont{{Kraiselburd}}},
  \bibinfo{author}{\bibfnamefont{F.~L.} \bibnamefont{{Castillo}}},
  \bibinfo{author}{\bibfnamefont{M.~E.} \bibnamefont{{Mosquera}}},
  \bibnamefont{and}
  \bibinfo{author}{\bibfnamefont{H.}~\bibnamefont{{Vucetich}}},
  \bibinfo{journal}{\prd} \textbf{\bibinfo{volume}{97}}, \bibinfo{eid}{043526}
  (\bibinfo{year}{2018}), \eprint{1801.08594}.

\bibitem[{\citenamefont{{Hart} and {Chluba}}(2020)}]{Hart:2019dxi}
\bibinfo{author}{\bibfnamefont{L.}~\bibnamefont{{Hart}}} \bibnamefont{and}
  \bibinfo{author}{\bibfnamefont{J.}~\bibnamefont{{Chluba}}},
  \bibinfo{journal}{\mnras} \textbf{\bibinfo{volume}{493}},
  \bibinfo{pages}{3255} (\bibinfo{year}{2020}), \eprint{1912.03986}.

\bibitem[{\citenamefont{{Planck Collaboration}
  et~al.}(2015)\citenamefont{{Planck Collaboration}, {Ade}, {Aghanim},
  {Arnaud}, {Ashdown}, {Aumont}, {Baccigalupi}, {Banday}, {Barreiro},
  {Battaner} et~al.}}]{Planck2015}
\bibinfo{author}{\bibnamefont{{Planck Collaboration}}},
  \bibinfo{author}{\bibfnamefont{P.~A.~R.} \bibnamefont{{Ade}}},
  \bibinfo{author}{\bibfnamefont{N.}~\bibnamefont{{Aghanim}}},
  \bibinfo{author}{\bibfnamefont{M.}~\bibnamefont{{Arnaud}}},
  \bibinfo{author}{\bibfnamefont{M.}~\bibnamefont{{Ashdown}}},
  \bibinfo{author}{\bibfnamefont{J.}~\bibnamefont{{Aumont}}},
  \bibinfo{author}{\bibfnamefont{C.}~\bibnamefont{{Baccigalupi}}},
  \bibinfo{author}{\bibfnamefont{A.~J.} \bibnamefont{{Banday}}},
  \bibinfo{author}{\bibfnamefont{R.~B.} \bibnamefont{{Barreiro}}},
  \bibinfo{author}{\bibfnamefont{E.}~\bibnamefont{{Battaner}}},
  \bibnamefont{et~al.}, \bibinfo{journal}{\aap} \textbf{\bibinfo{volume}{580}},
  \bibinfo{eid}{A22} (\bibinfo{year}{2015}), \eprint{1406.7482}.

\bibitem[{\citenamefont{{Hart} and {Chluba}}(2018)}]{2018MNRAS.474.1850H}
\bibinfo{author}{\bibfnamefont{L.}~\bibnamefont{{Hart}}} \bibnamefont{and}
  \bibinfo{author}{\bibfnamefont{J.}~\bibnamefont{{Chluba}}},
  \bibinfo{journal}{\mnras} \textbf{\bibinfo{volume}{474}},
  \bibinfo{pages}{1850} (\bibinfo{year}{2018}), \eprint{1705.03925}.

\bibitem[{\citenamefont{{Hinkley} et~al.}(2013)\citenamefont{{Hinkley},
  {Sherman}, {Phillips}, {Schioppo}, {Lemke}, {Beloy}, {Pizzocaro}, {Oates},
  and {Ludlow}}}]{Hinkley2013}
\bibinfo{author}{\bibfnamefont{N.}~\bibnamefont{{Hinkley}}},
  \bibinfo{author}{\bibfnamefont{J.~A.} \bibnamefont{{Sherman}}},
  \bibinfo{author}{\bibfnamefont{N.~B.} \bibnamefont{{Phillips}}},
  \bibinfo{author}{\bibfnamefont{M.}~\bibnamefont{{Schioppo}}},
  \bibinfo{author}{\bibfnamefont{N.~D.} \bibnamefont{{Lemke}}},
  \bibinfo{author}{\bibfnamefont{K.}~\bibnamefont{{Beloy}}},
  \bibinfo{author}{\bibfnamefont{M.}~\bibnamefont{{Pizzocaro}}},
  \bibinfo{author}{\bibfnamefont{C.~W.} \bibnamefont{{Oates}}},
  \bibnamefont{and} \bibinfo{author}{\bibfnamefont{A.~D.}
  \bibnamefont{{Ludlow}}}, \bibinfo{journal}{Science}
  \textbf{\bibinfo{volume}{341}}, \bibinfo{pages}{1215} (\bibinfo{year}{2013}),
  \eprint{1305.5869}.

\bibitem[{\citenamefont{{Leefer} et~al.}(2013)\citenamefont{{Leefer}, {Weber},
  {Cing{\"o}z}, {Torgerson}, and {Budker}}}]{2013PhRvL.111f0801L}
\bibinfo{author}{\bibfnamefont{N.}~\bibnamefont{{Leefer}}},
  \bibinfo{author}{\bibfnamefont{C.~T.~M.} \bibnamefont{{Weber}}},
  \bibinfo{author}{\bibfnamefont{A.}~\bibnamefont{{Cing{\"o}z}}},
  \bibinfo{author}{\bibfnamefont{J.~R.} \bibnamefont{{Torgerson}}},
  \bibnamefont{and} \bibinfo{author}{\bibfnamefont{D.}~\bibnamefont{{Budker}}},
  \bibinfo{journal}{\prl} \textbf{\bibinfo{volume}{111}}, \bibinfo{eid}{060801}
  (\bibinfo{year}{2013}), \eprint{1304.6940}.

\bibitem[{\citenamefont{Dijck}(2020)}]{Dijck:2020kfb}
\bibinfo{author}{\bibfnamefont{E.~A.} \bibnamefont{Dijck}}, Ph.D. thesis,
  \bibinfo{school}{Groningen U.} (\bibinfo{year}{2020}).

\bibitem[{\citenamefont{{Damour}
  et~al.}(2002{\natexlab{a}})\citenamefont{{Damour}, {Piazza}, and
  {Veneziano}}}]{damour1}
\bibinfo{author}{\bibfnamefont{T.}~\bibnamefont{{Damour}}},
  \bibinfo{author}{\bibfnamefont{F.}~\bibnamefont{{Piazza}}}, \bibnamefont{and}
  \bibinfo{author}{\bibfnamefont{G.}~\bibnamefont{{Veneziano}}},
  \bibinfo{journal}{\prd} \textbf{\bibinfo{volume}{66}}, \bibinfo{eid}{046007}
  (\bibinfo{year}{2002}{\natexlab{a}}), \eprint{hep-th/0205111}.

\bibitem[{\citenamefont{{Martins} and {Vacher}}(2019)}]{Martins2019}
\bibinfo{author}{\bibfnamefont{C.~J.~A.~P.} \bibnamefont{{Martins}}}
  \bibnamefont{and} \bibinfo{author}{\bibfnamefont{L.}~\bibnamefont{{Vacher}}},
  \bibinfo{journal}{\prd} \textbf{\bibinfo{volume}{100}}, \bibinfo{eid}{123514}
  (\bibinfo{year}{2019}), \eprint{1911.10821}.

\bibitem[{\citenamefont{{Martinelli} et~al.}(2018)\citenamefont{{Martinelli},
  {Calabrese}, and {Martins}}}]{Martins2018}
\bibinfo{author}{\bibfnamefont{M.}~\bibnamefont{{Martinelli}}},
  \bibinfo{author}{\bibfnamefont{E.}~\bibnamefont{{Calabrese}}},
  \bibnamefont{and} \bibinfo{author}{\bibfnamefont{C.~J.~A.~P.}
  \bibnamefont{{Martins}}}, in \emph{\bibinfo{booktitle}{Fourteenth Marcel
  Grossmann Meeting - MG14}}, edited by
  \bibinfo{editor}{\bibfnamefont{M.}~\bibnamefont{{Bianchi}}},
  \bibinfo{editor}{\bibfnamefont{R.~T.} \bibnamefont{{Jansen}}},
  \bibnamefont{and} \bibinfo{editor}{\bibfnamefont{R.}~\bibnamefont{{Ruffini}}}
  (\bibinfo{year}{2018}), pp. \bibinfo{pages}{3664--3669}.

\bibitem[{\citenamefont{{Martinelli} et~al.}(2015)\citenamefont{{Martinelli},
  {Calabrese}, and {Martins}}}]{Martins2015}
\bibinfo{author}{\bibfnamefont{M.}~\bibnamefont{{Martinelli}}},
  \bibinfo{author}{\bibfnamefont{E.}~\bibnamefont{{Calabrese}}},
  \bibnamefont{and} \bibinfo{author}{\bibfnamefont{C.~J.~A.~P.}
  \bibnamefont{{Martins}}}, \bibinfo{journal}{\jcap}
  \textbf{\bibinfo{volume}{2015}}, \bibinfo{eid}{030} (\bibinfo{year}{2015}),
  \eprint{1508.00765}.

\bibitem[{\citenamefont{{Bora} and {Desai}}(2021)}]{Kamal2021}
\bibinfo{author}{\bibfnamefont{K.}~\bibnamefont{{Bora}}} \bibnamefont{and}
  \bibinfo{author}{\bibfnamefont{S.}~\bibnamefont{{Desai}}},
  \bibinfo{journal}{\jcap} \textbf{\bibinfo{volume}{2021}}, \bibinfo{eid}{012}
  (\bibinfo{year}{2021}), \eprint{2008.10541}.

\bibitem[{\citenamefont{{Cola{\c{c}}o}
  et~al.}(2021{\natexlab{a}})\citenamefont{{Cola{\c{c}}o}, {Holanda}, and
  {Silva}}}]{leo1}
\bibinfo{author}{\bibfnamefont{L.~R.} \bibnamefont{{Cola{\c{c}}o}}},
  \bibinfo{author}{\bibfnamefont{R.~F.~L.} \bibnamefont{{Holanda}}},
  \bibnamefont{and} \bibinfo{author}{\bibfnamefont{R.}~\bibnamefont{{Silva}}},
  \bibinfo{journal}{European Physical Journal C} \textbf{\bibinfo{volume}{81}},
  \bibinfo{eid}{822} (\bibinfo{year}{2021}{\natexlab{a}}), \eprint{2004.08484}.

\bibitem[{\citenamefont{{Cola{\c{c}}o}
  et~al.}(2019)\citenamefont{{Cola{\c{c}}o}, {Holanda}, {Silva}, and
  {Alcaniz}}}]{Colaco2019}
\bibinfo{author}{\bibfnamefont{L.~R.} \bibnamefont{{Cola{\c{c}}o}}},
  \bibinfo{author}{\bibfnamefont{R.~F.~L.} \bibnamefont{{Holanda}}},
  \bibinfo{author}{\bibfnamefont{R.}~\bibnamefont{{Silva}}}, \bibnamefont{and}
  \bibinfo{author}{\bibfnamefont{J.~S.} \bibnamefont{{Alcaniz}}},
  \bibinfo{journal}{\jcap} \textbf{\bibinfo{volume}{2019}}, \bibinfo{eid}{014}
  (\bibinfo{year}{2019}), \eprint{1901.10947}.

\bibitem[{\citenamefont{{Holanda}
  et~al.}(2016{\natexlab{a}})\citenamefont{{Holanda}, {Landau}, {Alcaniz},
  {S{\'a}nchez G.}, and {Busti}}}]{Holanda2016JCAP}
\bibinfo{author}{\bibfnamefont{R.~F.~L.} \bibnamefont{{Holanda}}},
  \bibinfo{author}{\bibfnamefont{S.~J.} \bibnamefont{{Landau}}},
  \bibinfo{author}{\bibfnamefont{J.~S.} \bibnamefont{{Alcaniz}}},
  \bibinfo{author}{\bibfnamefont{I.~E.} \bibnamefont{{S{\'a}nchez G.}}},
  \bibnamefont{and} \bibinfo{author}{\bibfnamefont{V.~C.}
  \bibnamefont{{Busti}}}, \bibinfo{journal}{\jcap}
  \textbf{\bibinfo{volume}{2016}}, \bibinfo{eid}{047}
  (\bibinfo{year}{2016}{\natexlab{a}}), \eprint{1510.07240}.

\bibitem[{\citenamefont{{Holanda}
  et~al.}(2016{\natexlab{b}})\citenamefont{{Holanda}, {Busti}, {Cola{\c{c}}o},
  {Alcaniz}, and {Landau}}}]{Holanda2016JCAP2}
\bibinfo{author}{\bibfnamefont{R.~F.~L.} \bibnamefont{{Holanda}}},
  \bibinfo{author}{\bibfnamefont{V.~C.} \bibnamefont{{Busti}}},
  \bibinfo{author}{\bibfnamefont{L.~R.} \bibnamefont{{Cola{\c{c}}o}}},
  \bibinfo{author}{\bibfnamefont{J.~S.} \bibnamefont{{Alcaniz}}},
  \bibnamefont{and} \bibinfo{author}{\bibfnamefont{S.~J.}
  \bibnamefont{{Landau}}}, \bibinfo{journal}{\jcap}
  \textbf{\bibinfo{volume}{2016}}, \bibinfo{eid}{055}
  (\bibinfo{year}{2016}{\natexlab{b}}), \eprint{1605.02578}.

\bibitem[{\citenamefont{{Martins} et~al.}(2015)\citenamefont{{Martins},
  {Vielzeuf}, {Martinelli}, {Calabrese}, and {Pandolfi}}}]{Martins20152}
\bibinfo{author}{\bibfnamefont{C.~J.~A.~P.} \bibnamefont{{Martins}}},
  \bibinfo{author}{\bibfnamefont{P.~E.} \bibnamefont{{Vielzeuf}}},
  \bibinfo{author}{\bibfnamefont{M.}~\bibnamefont{{Martinelli}}},
  \bibinfo{author}{\bibfnamefont{E.}~\bibnamefont{{Calabrese}}},
  \bibnamefont{and}
  \bibinfo{author}{\bibfnamefont{S.}~\bibnamefont{{Pandolfi}}},
  \bibinfo{journal}{Physics Letters B} \textbf{\bibinfo{volume}{743}},
  \bibinfo{pages}{377} (\bibinfo{year}{2015}), \eprint{1503.05068}.

\bibitem[{\citenamefont{Collaboration et~al.}(2021)\citenamefont{Collaboration,
  the Virgo~Collaboration, and the
  KAGRA~Collaboration}}]{theligoscientificcollaboration2021gwtc3}
\bibinfo{author}{\bibfnamefont{T.~L.~S.} \bibnamefont{Collaboration}},
  \bibinfo{author}{\bibnamefont{the Virgo~Collaboration}}, \bibnamefont{and}
  \bibinfo{author}{\bibnamefont{the KAGRA~Collaboration}},
  \emph{\bibinfo{title}{Gwtc-3: Compact binary coalescences observed by ligo
  and virgo during the second part of the third observing run}}
  (\bibinfo{year}{2021}), \eprint{2111.03606}.

\bibitem[{\citenamefont{{Schutz}}(1986)}]{1986Natur.323..310S}
\bibinfo{author}{\bibfnamefont{B.~F.} \bibnamefont{{Schutz}}},
  \bibinfo{journal}{\nat} \textbf{\bibinfo{volume}{323}}, \bibinfo{pages}{310}
  (\bibinfo{year}{1986}).

\bibitem[{\citenamefont{Holz and Hughes}(2005)}]{Holz_2005}
\bibinfo{author}{\bibfnamefont{D.~E.} \bibnamefont{Holz}} \bibnamefont{and}
  \bibinfo{author}{\bibfnamefont{S.~A.} \bibnamefont{Hughes}},
  \bibinfo{journal}{The Astrophysical Journal} \textbf{\bibinfo{volume}{629}},
  \bibinfo{pages}{15} (\bibinfo{year}{2005}),
  \urlprefix\url{https://doi.org/10.1086%2F431341}.

\bibitem[{\citenamefont{Collaboration}(2017)}]{Abbott_2017}
\bibinfo{author}{\bibfnamefont{B.~T. L.~S.} \bibnamefont{Collaboration}},
  \bibinfo{journal}{Physical Review Letters} \textbf{\bibinfo{volume}{119}}
  (\bibinfo{year}{2017}),
  \urlprefix\url{https://doi.org/10.1103%2Fphysrevlett.119.161101}.

\bibitem[{\citenamefont{Abbott et~al.}(2017)}]{LIGOScientific:2017adf}
\bibinfo{author}{\bibfnamefont{B.~P.} \bibnamefont{Abbott}}
  \bibnamefont{et~al.} (\bibinfo{collaboration}{LIGO Scientific, Virgo, 1M2H,
  Dark Energy Camera GW-E, DES, DLT40, Las Cumbres Observatory, VINROUGE,
  MASTER}), \bibinfo{journal}{Nature} \textbf{\bibinfo{volume}{551}},
  \bibinfo{pages}{85} (\bibinfo{year}{2017}), \eprint{1710.05835}.

\bibitem[{\citenamefont{Kase and Tsujikawa}(2019)}]{Kase:2018aps}
\bibinfo{author}{\bibfnamefont{R.}~\bibnamefont{Kase}} \bibnamefont{and}
  \bibinfo{author}{\bibfnamefont{S.}~\bibnamefont{Tsujikawa}},
  \bibinfo{journal}{Int. J. Mod. Phys. D} \textbf{\bibinfo{volume}{28}},
  \bibinfo{pages}{1942005} (\bibinfo{year}{2019}), \eprint{1809.08735}.

\bibitem[{\citenamefont{Maggiore
  et~al.}(2020{\natexlab{a}})}]{Maggiore:2019uih}
\bibinfo{author}{\bibfnamefont{M.}~\bibnamefont{Maggiore}}
  \bibnamefont{et~al.}, \bibinfo{journal}{JCAP} \textbf{\bibinfo{volume}{03}},
  \bibinfo{pages}{050} (\bibinfo{year}{2020}{\natexlab{a}}),
  \eprint{1912.02622}.

\bibitem[{\citenamefont{Reitze et~al.}(2019)}]{Reitze:2019iox}
\bibinfo{author}{\bibfnamefont{D.}~\bibnamefont{Reitze}} \bibnamefont{et~al.},
  \bibinfo{journal}{Bull. Am. Astron. Soc.} \textbf{\bibinfo{volume}{51}},
  \bibinfo{pages}{035} (\bibinfo{year}{2019}), \eprint{1907.04833}.

\bibitem[{\citenamefont{Amaro-Seoane et~al.}(2017)}]{LISA:2017pwj}
\bibinfo{author}{\bibfnamefont{P.}~\bibnamefont{Amaro-Seoane}}
  \bibnamefont{et~al.} (\bibinfo{collaboration}{LISA}) (\bibinfo{year}{2017}),
  \eprint{1702.00786}.

\bibitem[{\citenamefont{Kawamura et~al.}(2021)}]{Kawamura:2020pcg}
\bibinfo{author}{\bibfnamefont{S.}~\bibnamefont{Kawamura}}
  \bibnamefont{et~al.}, \bibinfo{journal}{PTEP}
  \textbf{\bibinfo{volume}{2021}}, \bibinfo{pages}{05A105}
  (\bibinfo{year}{2021}), \eprint{2006.13545}.

\bibitem[{\citenamefont{Luo et~al.}(2016)}]{TianQin:2015yph}
\bibinfo{author}{\bibfnamefont{J.}~\bibnamefont{Luo}} \bibnamefont{et~al.}
  (\bibinfo{collaboration}{TianQin}), \bibinfo{journal}{Class. Quant. Grav.}
  \textbf{\bibinfo{volume}{33}}, \bibinfo{pages}{035010}
  (\bibinfo{year}{2016}), \eprint{1512.02076}.

\bibitem[{\citenamefont{Cai and Yang}(2017{\natexlab{a}})}]{Cai:2016sby}
\bibinfo{author}{\bibfnamefont{R.-G.} \bibnamefont{Cai}} \bibnamefont{and}
  \bibinfo{author}{\bibfnamefont{T.}~\bibnamefont{Yang}},
  \bibinfo{journal}{Phys. Rev. D} \textbf{\bibinfo{volume}{95}},
  \bibinfo{pages}{044024} (\bibinfo{year}{2017}{\natexlab{a}}),
  \eprint{1608.08008}.

\bibitem[{\citenamefont{Du et~al.}(2019)\citenamefont{Du, Yang, Xu, Pan, and
  Mota}}]{Du:2018tia}
\bibinfo{author}{\bibfnamefont{M.}~\bibnamefont{Du}},
  \bibinfo{author}{\bibfnamefont{W.}~\bibnamefont{Yang}},
  \bibinfo{author}{\bibfnamefont{L.}~\bibnamefont{Xu}},
  \bibinfo{author}{\bibfnamefont{S.}~\bibnamefont{Pan}}, \bibnamefont{and}
  \bibinfo{author}{\bibfnamefont{D.~F.} \bibnamefont{Mota}},
  \bibinfo{journal}{Phys. Rev. D} \textbf{\bibinfo{volume}{100}},
  \bibinfo{pages}{043535} (\bibinfo{year}{2019}), \eprint{1812.01440}.

\bibitem[{\citenamefont{Zhang et~al.}(2019)\citenamefont{Zhang, Wang, Zhang,
  and Zhang}}]{Zhang:2018byx}
\bibinfo{author}{\bibfnamefont{X.-N.} \bibnamefont{Zhang}},
  \bibinfo{author}{\bibfnamefont{L.-F.} \bibnamefont{Wang}},
  \bibinfo{author}{\bibfnamefont{J.-F.} \bibnamefont{Zhang}}, \bibnamefont{and}
  \bibinfo{author}{\bibfnamefont{X.}~\bibnamefont{Zhang}},
  \bibinfo{journal}{Phys. Rev. D} \textbf{\bibinfo{volume}{99}},
  \bibinfo{pages}{063510} (\bibinfo{year}{2019}), \eprint{1804.08379}.

\bibitem[{\citenamefont{Yang et~al.}(2019)\citenamefont{Yang, Vagnozzi,
  Di~Valentino, Nunes, Pan, and Mota}}]{Yang:2019vni}
\bibinfo{author}{\bibfnamefont{W.}~\bibnamefont{Yang}},
  \bibinfo{author}{\bibfnamefont{S.}~\bibnamefont{Vagnozzi}},
  \bibinfo{author}{\bibfnamefont{E.}~\bibnamefont{Di~Valentino}},
  \bibinfo{author}{\bibfnamefont{R.~C.} \bibnamefont{Nunes}},
  \bibinfo{author}{\bibfnamefont{S.}~\bibnamefont{Pan}}, \bibnamefont{and}
  \bibinfo{author}{\bibfnamefont{D.~F.} \bibnamefont{Mota}},
  \bibinfo{journal}{JCAP} \textbf{\bibinfo{volume}{07}}, \bibinfo{pages}{037}
  (\bibinfo{year}{2019}), \eprint{1905.08286}.

\bibitem[{\citenamefont{Fu et~al.}(2019)\citenamefont{Fu, Zhou, and
  Chen}}]{Fu:2019oll}
\bibinfo{author}{\bibfnamefont{X.}~\bibnamefont{Fu}},
  \bibinfo{author}{\bibfnamefont{L.}~\bibnamefont{Zhou}}, \bibnamefont{and}
  \bibinfo{author}{\bibfnamefont{J.}~\bibnamefont{Chen}},
  \bibinfo{journal}{Phys. Rev. D} \textbf{\bibinfo{volume}{99}},
  \bibinfo{pages}{083523} (\bibinfo{year}{2019}), \eprint{1903.09913}.

\bibitem[{\citenamefont{Cai et~al.}(2018)\citenamefont{Cai, Liu, Liu, Wang, and
  Yang}}]{Cai:2017aea}
\bibinfo{author}{\bibfnamefont{R.-G.} \bibnamefont{Cai}},
  \bibinfo{author}{\bibfnamefont{T.-B.} \bibnamefont{Liu}},
  \bibinfo{author}{\bibfnamefont{X.-W.} \bibnamefont{Liu}},
  \bibinfo{author}{\bibfnamefont{S.-J.} \bibnamefont{Wang}}, \bibnamefont{and}
  \bibinfo{author}{\bibfnamefont{T.}~\bibnamefont{Yang}},
  \bibinfo{journal}{Phys. Rev. D} \textbf{\bibinfo{volume}{97}},
  \bibinfo{pages}{103005} (\bibinfo{year}{2018}), \eprint{1712.00952}.

\bibitem[{\citenamefont{Allahyari et~al.}(2022)\citenamefont{Allahyari, Nunes,
  and Mota}}]{Allahyari:2021enz}
\bibinfo{author}{\bibfnamefont{A.}~\bibnamefont{Allahyari}},
  \bibinfo{author}{\bibfnamefont{R.~C.} \bibnamefont{Nunes}}, \bibnamefont{and}
  \bibinfo{author}{\bibfnamefont{D.~F.} \bibnamefont{Mota}},
  \bibinfo{journal}{Mon. Not. Roy. Astron. Soc.}
  \textbf{\bibinfo{volume}{514}}, \bibinfo{pages}{1274} (\bibinfo{year}{2022}),
  \eprint{2110.07634}.

\bibitem[{\citenamefont{Belgacem et~al.}(2018)\citenamefont{Belgacem, Dirian,
  Foffa, and Maggiore}}]{Belgacem:2017ihm}
\bibinfo{author}{\bibfnamefont{E.}~\bibnamefont{Belgacem}},
  \bibinfo{author}{\bibfnamefont{Y.}~\bibnamefont{Dirian}},
  \bibinfo{author}{\bibfnamefont{S.}~\bibnamefont{Foffa}}, \bibnamefont{and}
  \bibinfo{author}{\bibfnamefont{M.}~\bibnamefont{Maggiore}},
  \bibinfo{journal}{Phys. Rev. D} \textbf{\bibinfo{volume}{97}},
  \bibinfo{pages}{104066} (\bibinfo{year}{2018}), \eprint{1712.08108}.

\bibitem[{\citenamefont{D'Agostino and Nunes}(2019)}]{DAgostino:2019hvh}
\bibinfo{author}{\bibfnamefont{R.}~\bibnamefont{D'Agostino}} \bibnamefont{and}
  \bibinfo{author}{\bibfnamefont{R.~C.} \bibnamefont{Nunes}},
  \bibinfo{journal}{Phys. Rev. D} \textbf{\bibinfo{volume}{100}},
  \bibinfo{pages}{044041} (\bibinfo{year}{2019}), \eprint{1907.05516}.

\bibitem[{\citenamefont{Nishizawa and Arai}(2019)}]{Nishizawa:2019rra}
\bibinfo{author}{\bibfnamefont{A.}~\bibnamefont{Nishizawa}} \bibnamefont{and}
  \bibinfo{author}{\bibfnamefont{S.}~\bibnamefont{Arai}},
  \bibinfo{journal}{Phys. Rev. D} \textbf{\bibinfo{volume}{99}},
  \bibinfo{pages}{104038} (\bibinfo{year}{2019}), \eprint{1901.08249}.

\bibitem[{\citenamefont{Bonilla et~al.}(2020)\citenamefont{Bonilla, D'Agostino,
  Nunes, and de~Araujo}}]{Bonilla:2019mbm}
\bibinfo{author}{\bibfnamefont{A.}~\bibnamefont{Bonilla}},
  \bibinfo{author}{\bibfnamefont{R.}~\bibnamefont{D'Agostino}},
  \bibinfo{author}{\bibfnamefont{R.~C.} \bibnamefont{Nunes}}, \bibnamefont{and}
  \bibinfo{author}{\bibfnamefont{J.~C.~N.} \bibnamefont{de~Araujo}},
  \bibinfo{journal}{JCAP} \textbf{\bibinfo{volume}{03}}, \bibinfo{pages}{015}
  (\bibinfo{year}{2020}), \eprint{1910.05631}.

\bibitem[{\citenamefont{Odintsov et~al.}(2022)\citenamefont{Odintsov,
  Oikonomou, and Myrzakulov}}]{Odintsov:2022cbm}
\bibinfo{author}{\bibfnamefont{S.~D.} \bibnamefont{Odintsov}},
  \bibinfo{author}{\bibfnamefont{V.~K.} \bibnamefont{Oikonomou}},
  \bibnamefont{and}
  \bibinfo{author}{\bibfnamefont{R.}~\bibnamefont{Myrzakulov}},
  \bibinfo{journal}{Symmetry} \textbf{\bibinfo{volume}{14}},
  \bibinfo{pages}{729} (\bibinfo{year}{2022}), \eprint{2204.00876}.

\bibitem[{\citenamefont{Cai and Yang}(2021)}]{Cai:2021ooo}
\bibinfo{author}{\bibfnamefont{R.-G.} \bibnamefont{Cai}} \bibnamefont{and}
  \bibinfo{author}{\bibfnamefont{T.}~\bibnamefont{Yang}},
  \bibinfo{journal}{JCAP} \textbf{\bibinfo{volume}{12}}, \bibinfo{pages}{017}
  (\bibinfo{year}{2021}), \eprint{2107.13919}.

\bibitem[{\citenamefont{Matos et~al.}(2021)\citenamefont{Matos, Calv\~ao, and
  Waga}}]{Matos:2021qne}
\bibinfo{author}{\bibfnamefont{I.~S.} \bibnamefont{Matos}},
  \bibinfo{author}{\bibfnamefont{M.~O.} \bibnamefont{Calv\~ao}},
  \bibnamefont{and} \bibinfo{author}{\bibfnamefont{I.}~\bibnamefont{Waga}},
  \bibinfo{journal}{Phys. Rev. D} \textbf{\bibinfo{volume}{103}},
  \bibinfo{pages}{104059} (\bibinfo{year}{2021}), \eprint{2104.10305}.

\bibitem[{\citenamefont{Jiang and Yagi}(2021)}]{Jiang:2021mpd}
\bibinfo{author}{\bibfnamefont{N.}~\bibnamefont{Jiang}} \bibnamefont{and}
  \bibinfo{author}{\bibfnamefont{K.}~\bibnamefont{Yagi}},
  \bibinfo{journal}{Phys. Rev. D} \textbf{\bibinfo{volume}{103}},
  \bibinfo{pages}{124047} (\bibinfo{year}{2021}), \eprint{2104.04442}.

\bibitem[{\citenamefont{Pan et~al.}(2021)\citenamefont{Pan, He, Qi, Li, Cao,
  Liu, and Wang}}]{Pan:2021tpk}
\bibinfo{author}{\bibfnamefont{Y.}~\bibnamefont{Pan}},
  \bibinfo{author}{\bibfnamefont{Y.}~\bibnamefont{He}},
  \bibinfo{author}{\bibfnamefont{J.}~\bibnamefont{Qi}},
  \bibinfo{author}{\bibfnamefont{J.}~\bibnamefont{Li}},
  \bibinfo{author}{\bibfnamefont{S.}~\bibnamefont{Cao}},
  \bibinfo{author}{\bibfnamefont{T.}~\bibnamefont{Liu}}, \bibnamefont{and}
  \bibinfo{author}{\bibfnamefont{J.}~\bibnamefont{Wang}},
  \bibinfo{journal}{Astrophys. J.} \textbf{\bibinfo{volume}{911}},
  \bibinfo{pages}{135} (\bibinfo{year}{2021}), \eprint{2103.05212}.

\bibitem[{\citenamefont{Tasinato et~al.}(2021)\citenamefont{Tasinato,
  Garoffolo, Bertacca, and Matarrese}}]{Tasinato:2021wol}
\bibinfo{author}{\bibfnamefont{G.}~\bibnamefont{Tasinato}},
  \bibinfo{author}{\bibfnamefont{A.}~\bibnamefont{Garoffolo}},
  \bibinfo{author}{\bibfnamefont{D.}~\bibnamefont{Bertacca}}, \bibnamefont{and}
  \bibinfo{author}{\bibfnamefont{S.}~\bibnamefont{Matarrese}},
  \bibinfo{journal}{JCAP} \textbf{\bibinfo{volume}{06}}, \bibinfo{pages}{050}
  (\bibinfo{year}{2021}), \eprint{2103.00155}.

\bibitem[{\citenamefont{Bonilla et~al.}(2022)\citenamefont{Bonilla, Kumar,
  Nunes, and Pan}}]{Bonilla:2021dql}
\bibinfo{author}{\bibfnamefont{A.}~\bibnamefont{Bonilla}},
  \bibinfo{author}{\bibfnamefont{S.}~\bibnamefont{Kumar}},
  \bibinfo{author}{\bibfnamefont{R.~C.} \bibnamefont{Nunes}}, \bibnamefont{and}
  \bibinfo{author}{\bibfnamefont{S.}~\bibnamefont{Pan}}, \bibinfo{journal}{Mon.
  Not. Roy. Astron. Soc.} \textbf{\bibinfo{volume}{512}}, \bibinfo{pages}{4231}
  (\bibinfo{year}{2022}), \eprint{2102.06149}.

\bibitem[{\citenamefont{Mukherjee et~al.}(2021)\citenamefont{Mukherjee,
  Wandelt, and Silk}}]{Mukherjee:2020mha}
\bibinfo{author}{\bibfnamefont{S.}~\bibnamefont{Mukherjee}},
  \bibinfo{author}{\bibfnamefont{B.~D.} \bibnamefont{Wandelt}},
  \bibnamefont{and} \bibinfo{author}{\bibfnamefont{J.}~\bibnamefont{Silk}},
  \bibinfo{journal}{Mon. Not. Roy. Astron. Soc.}
  \textbf{\bibinfo{volume}{502}}, \bibinfo{pages}{1136} (\bibinfo{year}{2021}),
  \eprint{2012.15316}.

\bibitem[{\citenamefont{Kalomenopoulos
  et~al.}(2021)\citenamefont{Kalomenopoulos, Khochfar, Gair, and
  Arai}}]{Kalomenopoulos:2020klp}
\bibinfo{author}{\bibfnamefont{M.}~\bibnamefont{Kalomenopoulos}},
  \bibinfo{author}{\bibfnamefont{S.}~\bibnamefont{Khochfar}},
  \bibinfo{author}{\bibfnamefont{J.}~\bibnamefont{Gair}}, \bibnamefont{and}
  \bibinfo{author}{\bibfnamefont{S.}~\bibnamefont{Arai}},
  \bibinfo{journal}{Mon. Not. Roy. Astron. Soc.}
  \textbf{\bibinfo{volume}{503}}, \bibinfo{pages}{3179} (\bibinfo{year}{2021}),
  \eprint{2007.15020}.

\bibitem[{\citenamefont{Baker and Harrison}(2021)}]{Baker:2020apq}
\bibinfo{author}{\bibfnamefont{T.}~\bibnamefont{Baker}} \bibnamefont{and}
  \bibinfo{author}{\bibfnamefont{I.}~\bibnamefont{Harrison}},
  \bibinfo{journal}{JCAP} \textbf{\bibinfo{volume}{01}}, \bibinfo{pages}{068}
  (\bibinfo{year}{2021}), \eprint{2007.13791}.

\bibitem[{\citenamefont{Mastrogiovanni
  et~al.}(2020)\citenamefont{Mastrogiovanni, Steer, and
  Barsuglia}}]{Mastrogiovanni:2020gua}
\bibinfo{author}{\bibfnamefont{S.}~\bibnamefont{Mastrogiovanni}},
  \bibinfo{author}{\bibfnamefont{D.}~\bibnamefont{Steer}}, \bibnamefont{and}
  \bibinfo{author}{\bibfnamefont{M.}~\bibnamefont{Barsuglia}},
  \bibinfo{journal}{Phys. Rev. D} \textbf{\bibinfo{volume}{102}},
  \bibinfo{pages}{044009} (\bibinfo{year}{2020}), \eprint{2004.01632}.

\bibitem[{\citenamefont{Belgacem et~al.}(2020)\citenamefont{Belgacem, Foffa,
  Maggiore, and Yang}}]{Belgacem:2019zzu}
\bibinfo{author}{\bibfnamefont{E.}~\bibnamefont{Belgacem}},
  \bibinfo{author}{\bibfnamefont{S.}~\bibnamefont{Foffa}},
  \bibinfo{author}{\bibfnamefont{M.}~\bibnamefont{Maggiore}}, \bibnamefont{and}
  \bibinfo{author}{\bibfnamefont{T.}~\bibnamefont{Yang}},
  \bibinfo{journal}{Phys. Rev. D} \textbf{\bibinfo{volume}{101}},
  \bibinfo{pages}{063505} (\bibinfo{year}{2020}), \eprint{1911.11497}.

\bibitem[{\citenamefont{Nunes et~al.}(2019)\citenamefont{Nunes, Alves, and
  de~Araujo}}]{Nunes:2019bjq}
\bibinfo{author}{\bibfnamefont{R.~C.} \bibnamefont{Nunes}},
  \bibinfo{author}{\bibfnamefont{M.~E.~S.} \bibnamefont{Alves}},
  \bibnamefont{and} \bibinfo{author}{\bibfnamefont{J.~C.~N.}
  \bibnamefont{de~Araujo}}, \bibinfo{journal}{Phys. Rev. D}
  \textbf{\bibinfo{volume}{100}}, \bibinfo{pages}{064012}
  (\bibinfo{year}{2019}), \eprint{1905.03237}.

\bibitem[{\citenamefont{Harry and Noller}(2022)}]{Harry:2022zey}
\bibinfo{author}{\bibfnamefont{I.}~\bibnamefont{Harry}} \bibnamefont{and}
  \bibinfo{author}{\bibfnamefont{J.}~\bibnamefont{Noller}}
  (\bibinfo{year}{2022}), \eprint{2207.10096}.

\bibitem[{\citenamefont{Ezquiaga et~al.}(2021)\citenamefont{Ezquiaga, Hu,
  Lagos, and Lin}}]{Ezquiaga:2021ler}
\bibinfo{author}{\bibfnamefont{J.~M.} \bibnamefont{Ezquiaga}},
  \bibinfo{author}{\bibfnamefont{W.}~\bibnamefont{Hu}},
  \bibinfo{author}{\bibfnamefont{M.}~\bibnamefont{Lagos}}, \bibnamefont{and}
  \bibinfo{author}{\bibfnamefont{M.-X.} \bibnamefont{Lin}},
  \bibinfo{journal}{JCAP} \textbf{\bibinfo{volume}{11}}, \bibinfo{pages}{048}
  (\bibinfo{year}{2021}), \eprint{2108.10872}.

\bibitem[{\citenamefont{{Cheung} et~al.}(2021)\citenamefont{{Cheung}, {Gais},
  {Hannuksela}, and {Li}}}]{2021MNRAS.503.3326C}
\bibinfo{author}{\bibfnamefont{M.~H.~Y.} \bibnamefont{{Cheung}}},
  \bibinfo{author}{\bibfnamefont{J.}~\bibnamefont{{Gais}}},
  \bibinfo{author}{\bibfnamefont{O.~A.} \bibnamefont{{Hannuksela}}},
  \bibnamefont{and} \bibinfo{author}{\bibfnamefont{T.~G.~F.}
  \bibnamefont{{Li}}}, \bibinfo{journal}{\mnras}
  \textbf{\bibinfo{volume}{503}}, \bibinfo{pages}{3326} (\bibinfo{year}{2021}),
  \eprint{2012.07800}.

\bibitem[{\citenamefont{{Yeung} et~al.}(2021)\citenamefont{{Yeung}, {Cheung},
  {Gais}, {Hannuksela}, and {Li}}}]{2021arXiv211207635Y}
\bibinfo{author}{\bibfnamefont{S.~M.~C.} \bibnamefont{{Yeung}}},
  \bibinfo{author}{\bibfnamefont{M.~H.~Y.} \bibnamefont{{Cheung}}},
  \bibinfo{author}{\bibfnamefont{J.~A.~J.} \bibnamefont{{Gais}}},
  \bibinfo{author}{\bibfnamefont{O.~A.} \bibnamefont{{Hannuksela}}},
  \bibnamefont{and} \bibinfo{author}{\bibfnamefont{T.~G.~F.}
  \bibnamefont{{Li}}}, \bibinfo{journal}{arXiv e-prints}
  \bibinfo{eid}{arXiv:2112.07635} (\bibinfo{year}{2021}), \eprint{2112.07635}.

\bibitem[{\citenamefont{{Leung} et~al.}(2023)\citenamefont{{Leung}, {Jow},
  {Saha}, {Dai}, {Oguri}, and {Koopmans}}}]{2023arXiv230401202L}
\bibinfo{author}{\bibfnamefont{C.}~\bibnamefont{{Leung}}},
  \bibinfo{author}{\bibfnamefont{D.}~\bibnamefont{{Jow}}},
  \bibinfo{author}{\bibfnamefont{P.}~\bibnamefont{{Saha}}},
  \bibinfo{author}{\bibfnamefont{L.}~\bibnamefont{{Dai}}},
  \bibinfo{author}{\bibfnamefont{M.}~\bibnamefont{{Oguri}}}, \bibnamefont{and}
  \bibinfo{author}{\bibfnamefont{L.~V.~E.} \bibnamefont{{Koopmans}}},
  \bibinfo{journal}{arXiv e-prints} \bibinfo{eid}{arXiv:2304.01202}
  (\bibinfo{year}{2023}), \eprint{2304.01202}.

\bibitem[{\citenamefont{Efstathiou}(2021)}]{efstathiou2021h0}
\bibinfo{author}{\bibfnamefont{G.}~\bibnamefont{Efstathiou}},
  \emph{\bibinfo{title}{To h0 or not to h0?}} (\bibinfo{year}{2021}),
  \eprint{2103.08723}.

\bibitem[{\citenamefont{Camarena and Marra}(2021)}]{Camarena_2021}
\bibinfo{author}{\bibfnamefont{D.}~\bibnamefont{Camarena}} \bibnamefont{and}
  \bibinfo{author}{\bibfnamefont{V.}~\bibnamefont{Marra}},
  \bibinfo{journal}{Monthly Notices of the Royal Astronomical Society}
  \textbf{\bibinfo{volume}{504}}, \bibinfo{pages}{5164–5171}
  (\bibinfo{year}{2021}), ISSN \bibinfo{issn}{1365-2966},
  \urlprefix\url{http://dx.doi.org/10.1093/mnras/stab1200}.

\bibitem[{\citenamefont{Nunes and Di~Valentino}(2021)}]{Nunes2021c}
\bibinfo{author}{\bibfnamefont{R.~C.} \bibnamefont{Nunes}} \bibnamefont{and}
  \bibinfo{author}{\bibfnamefont{E.}~\bibnamefont{Di~Valentino}},
  \bibinfo{journal}{Physical Review D} \textbf{\bibinfo{volume}{104}}
  (\bibinfo{year}{2021}), ISSN \bibinfo{issn}{2470-0029},
  \urlprefix\url{http://dx.doi.org/10.1103/PhysRevD.104.063529}.

\bibitem[{\citenamefont{{Cao} et~al.}(2015)\citenamefont{{Cao}, {Biesiada},
  {Gavazzi}, {Pi{\'o}rkowska}, and {Zhu}}}]{cao2015}
\bibinfo{author}{\bibfnamefont{S.}~\bibnamefont{{Cao}}},
  \bibinfo{author}{\bibfnamefont{M.}~\bibnamefont{{Biesiada}}},
  \bibinfo{author}{\bibfnamefont{R.}~\bibnamefont{{Gavazzi}}},
  \bibinfo{author}{\bibfnamefont{A.}~\bibnamefont{{Pi{\'o}rkowska}}},
  \bibnamefont{and} \bibinfo{author}{\bibfnamefont{Z.-H.} \bibnamefont{{Zhu}}},
  \bibinfo{journal}{\apj} \textbf{\bibinfo{volume}{806}}, \bibinfo{eid}{185}
  (\bibinfo{year}{2015}), \eprint{1509.07649}.

\bibitem[{\citenamefont{{Refsdal}}(1964)}]{Refsdal}
\bibinfo{author}{\bibfnamefont{S.}~\bibnamefont{{Refsdal}}},
  \bibinfo{journal}{Mont. Not. Royal. Astron. Soc.}
  \textbf{\bibinfo{volume}{128}}, \bibinfo{pages}{307} (\bibinfo{year}{1964}).

\bibitem[{\citenamefont{{Suyu} et~al.}(2010)\citenamefont{{Suyu}, {Marshall},
  {Auger}, {Hilbert}, {Blandford}, {Koopmans}, {Fassnacht}, and {Treu}}}]{Suyu}
\bibinfo{author}{\bibfnamefont{S.~H.} \bibnamefont{{Suyu}}},
  \bibinfo{author}{\bibfnamefont{P.~J.} \bibnamefont{{Marshall}}},
  \bibinfo{author}{\bibfnamefont{M.~W.} \bibnamefont{{Auger}}},
  \bibinfo{author}{\bibfnamefont{S.}~\bibnamefont{{Hilbert}}},
  \bibinfo{author}{\bibfnamefont{R.~D.} \bibnamefont{{Blandford}}},
  \bibinfo{author}{\bibfnamefont{L.~V.~E.} \bibnamefont{{Koopmans}}},
  \bibinfo{author}{\bibfnamefont{C.~D.} \bibnamefont{{Fassnacht}}},
  \bibnamefont{and} \bibinfo{author}{\bibfnamefont{T.}~\bibnamefont{{Treu}}},
  \bibinfo{journal}{\apj} \textbf{\bibinfo{volume}{711}}, \bibinfo{pages}{201}
  (\bibinfo{year}{2010}), \eprint{0910.2773}.

\bibitem[{\citenamefont{{Wong} et~al.}(2020)\citenamefont{{Wong}, {Suyu},
  {Chen}, {Rusu}, {Millon}, {Sluse}, {Bonvin}, {Fassnacht}, {Taubenberger},
  {Auger} et~al.}}]{H0LiCOW}
\bibinfo{author}{\bibfnamefont{K.~C.} \bibnamefont{{Wong}}},
  \bibinfo{author}{\bibfnamefont{S.~H.} \bibnamefont{{Suyu}}},
  \bibinfo{author}{\bibfnamefont{G.~C.~F.} \bibnamefont{{Chen}}},
  \bibinfo{author}{\bibfnamefont{C.~E.} \bibnamefont{{Rusu}}},
  \bibinfo{author}{\bibfnamefont{M.}~\bibnamefont{{Millon}}},
  \bibinfo{author}{\bibfnamefont{D.}~\bibnamefont{{Sluse}}},
  \bibinfo{author}{\bibfnamefont{V.}~\bibnamefont{{Bonvin}}},
  \bibinfo{author}{\bibfnamefont{C.~D.} \bibnamefont{{Fassnacht}}},
  \bibinfo{author}{\bibfnamefont{S.}~\bibnamefont{{Taubenberger}}},
  \bibinfo{author}{\bibfnamefont{M.~W.} \bibnamefont{{Auger}}},
  \bibnamefont{et~al.}, \bibinfo{journal}{Mont. Not. Royal. Astron. Soc.}
  \textbf{\bibinfo{volume}{498}}, \bibinfo{pages}{1420} (\bibinfo{year}{2020}),
  \eprint{1907.04869}.

\bibitem[{\citenamefont{{Shajib} et~al.}(2020)\citenamefont{{Shajib}, {Birrer},
  {Treu}, {Agnello}, {Buckley-Geer}, {Chan}, {Christensen}, {Lemon}, {Lin},
  {Millon} et~al.}}]{Shajib}
\bibinfo{author}{\bibfnamefont{A.~J.} \bibnamefont{{Shajib}}},
  \bibinfo{author}{\bibfnamefont{S.}~\bibnamefont{{Birrer}}},
  \bibinfo{author}{\bibfnamefont{T.}~\bibnamefont{{Treu}}},
  \bibinfo{author}{\bibfnamefont{A.}~\bibnamefont{{Agnello}}},
  \bibinfo{author}{\bibfnamefont{E.~J.} \bibnamefont{{Buckley-Geer}}},
  \bibinfo{author}{\bibfnamefont{J.~H.~H.} \bibnamefont{{Chan}}},
  \bibinfo{author}{\bibfnamefont{L.}~\bibnamefont{{Christensen}}},
  \bibinfo{author}{\bibfnamefont{C.}~\bibnamefont{{Lemon}}},
  \bibinfo{author}{\bibfnamefont{H.}~\bibnamefont{{Lin}}},
  \bibinfo{author}{\bibfnamefont{M.}~\bibnamefont{{Millon}}},
  \bibnamefont{et~al.}, \bibinfo{journal}{Mont. Not. Royal. Astron. Soc.}
  \textbf{\bibinfo{volume}{494}}, \bibinfo{pages}{6072} (\bibinfo{year}{2020}),
  \eprint{1910.06306}.

\bibitem[{\citenamefont{{Birrer} et~al.}(2020)\citenamefont{{Birrer}, {Shajib},
  {Galan}, {Millon}, {Treu}, {Agnello}, {Auger}, {Chen}, {Christensen},
  {Collett} et~al.}}]{Birrer}
\bibinfo{author}{\bibfnamefont{S.}~\bibnamefont{{Birrer}}},
  \bibinfo{author}{\bibfnamefont{A.~J.} \bibnamefont{{Shajib}}},
  \bibinfo{author}{\bibfnamefont{A.}~\bibnamefont{{Galan}}},
  \bibinfo{author}{\bibfnamefont{M.}~\bibnamefont{{Millon}}},
  \bibinfo{author}{\bibfnamefont{T.}~\bibnamefont{{Treu}}},
  \bibinfo{author}{\bibfnamefont{A.}~\bibnamefont{{Agnello}}},
  \bibinfo{author}{\bibfnamefont{M.}~\bibnamefont{{Auger}}},
  \bibinfo{author}{\bibfnamefont{G.~C.~F.} \bibnamefont{{Chen}}},
  \bibinfo{author}{\bibfnamefont{L.}~\bibnamefont{{Christensen}}},
  \bibinfo{author}{\bibfnamefont{T.}~\bibnamefont{{Collett}}},
  \bibnamefont{et~al.}, \bibinfo{journal}{\aap} \textbf{\bibinfo{volume}{643}},
  \bibinfo{eid}{A165} (\bibinfo{year}{2020}), \eprint{2007.02941}.

\bibitem[{\citenamefont{{Cola{\c{c}}o}
  et~al.}(2021{\natexlab{b}})\citenamefont{{Cola{\c{c}}o}, {Gonzalez}, and
  {Holanda}}}]{leo2}
\bibinfo{author}{\bibfnamefont{L.~R.} \bibnamefont{{Cola{\c{c}}o}}},
  \bibinfo{author}{\bibfnamefont{J.~E.} \bibnamefont{{Gonzalez}}},
  \bibnamefont{and} \bibinfo{author}{\bibfnamefont{R.~F.~L.}
  \bibnamefont{{Holanda}}}, \bibinfo{journal}{European Physical Journal C}
  \textbf{\bibinfo{volume}{81}}, \bibinfo{eid}{533}
  (\bibinfo{year}{2021}{\natexlab{b}}), \eprint{2010.04021}.

\bibitem[{\citenamefont{{Holanda} et~al.}(2017)\citenamefont{{Holanda},
  {Busti}, {Lima}, and {Alcaniz}}}]{holg}
\bibinfo{author}{\bibfnamefont{R.~F.~L.} \bibnamefont{{Holanda}}},
  \bibinfo{author}{\bibfnamefont{V.~C.} \bibnamefont{{Busti}}},
  \bibinfo{author}{\bibfnamefont{F.~S.} \bibnamefont{{Lima}}},
  \bibnamefont{and} \bibinfo{author}{\bibfnamefont{J.~S.}
  \bibnamefont{{Alcaniz}}}, \bibinfo{journal}{\jcap}
  \textbf{\bibinfo{volume}{2017}}, \bibinfo{eid}{039} (\bibinfo{year}{2017}),
  \eprint{1611.09426}.

\bibitem[{\citenamefont{{Koopmans} et~al.}(2009)\citenamefont{{Koopmans},
  {Bolton}, {Treu}, {Czoske}, {Auger}, {Barnab{\`e}}, {Vegetti}, {Gavazzi},
  {Moustakas}, and {Burles}}}]{Koopmans}
\bibinfo{author}{\bibfnamefont{L.~V.~E.} \bibnamefont{{Koopmans}}},
  \bibinfo{author}{\bibfnamefont{A.}~\bibnamefont{{Bolton}}},
  \bibinfo{author}{\bibfnamefont{T.}~\bibnamefont{{Treu}}},
  \bibinfo{author}{\bibfnamefont{O.}~\bibnamefont{{Czoske}}},
  \bibinfo{author}{\bibfnamefont{M.~W.} \bibnamefont{{Auger}}},
  \bibinfo{author}{\bibfnamefont{M.}~\bibnamefont{{Barnab{\`e}}}},
  \bibinfo{author}{\bibfnamefont{S.}~\bibnamefont{{Vegetti}}},
  \bibinfo{author}{\bibfnamefont{R.}~\bibnamefont{{Gavazzi}}},
  \bibinfo{author}{\bibfnamefont{L.~A.} \bibnamefont{{Moustakas}}},
  \bibnamefont{and} \bibinfo{author}{\bibfnamefont{S.}~\bibnamefont{{Burles}}},
  \bibinfo{journal}{\apjl} \textbf{\bibinfo{volume}{703}}, \bibinfo{pages}{L51}
  (\bibinfo{year}{2009}), \eprint{0906.1349}.

\bibitem[{\citenamefont{{Auger} et~al.}(2010)\citenamefont{{Auger}, {Treu},
  {Bolton}, {Gavazzi}, {Koopmans}, {Marshall}, {Moustakas}, and
  {Burles}}}]{Auger}
\bibinfo{author}{\bibfnamefont{M.~W.} \bibnamefont{{Auger}}},
  \bibinfo{author}{\bibfnamefont{T.}~\bibnamefont{{Treu}}},
  \bibinfo{author}{\bibfnamefont{A.~S.} \bibnamefont{{Bolton}}},
  \bibinfo{author}{\bibfnamefont{R.}~\bibnamefont{{Gavazzi}}},
  \bibinfo{author}{\bibfnamefont{L.~V.~E.} \bibnamefont{{Koopmans}}},
  \bibinfo{author}{\bibfnamefont{P.~J.} \bibnamefont{{Marshall}}},
  \bibinfo{author}{\bibfnamefont{L.~A.} \bibnamefont{{Moustakas}}},
  \bibnamefont{and} \bibinfo{author}{\bibfnamefont{S.}~\bibnamefont{{Burles}}},
  \bibinfo{journal}{\apj} \textbf{\bibinfo{volume}{724}}, \bibinfo{pages}{511}
  (\bibinfo{year}{2010}), \eprint{1007.2880}.

\bibitem[{\citenamefont{{Barnab{\`e}} et~al.}(2011)\citenamefont{{Barnab{\`e}},
  {Czoske}, {Koopmans}, {Treu}, and {Bolton}}}]{Barnab}
\bibinfo{author}{\bibfnamefont{M.}~\bibnamefont{{Barnab{\`e}}}},
  \bibinfo{author}{\bibfnamefont{O.}~\bibnamefont{{Czoske}}},
  \bibinfo{author}{\bibfnamefont{L.~V.~E.} \bibnamefont{{Koopmans}}},
  \bibinfo{author}{\bibfnamefont{T.}~\bibnamefont{{Treu}}}, \bibnamefont{and}
  \bibinfo{author}{\bibfnamefont{A.~S.} \bibnamefont{{Bolton}}},
  \bibinfo{journal}{Mont. Not. Royal. Astron. Soc.}
  \textbf{\bibinfo{volume}{415}}, \bibinfo{pages}{2215} (\bibinfo{year}{2011}),
  \eprint{1102.2261}.

\bibitem[{\citenamefont{{Sonnenfeld} et~al.}(2013)\citenamefont{{Sonnenfeld},
  {Treu}, {Gavazzi}, {Suyu}, {Marshall}, {Auger}, and {Nipoti}}}]{Sonnenfeld}
\bibinfo{author}{\bibfnamefont{A.}~\bibnamefont{{Sonnenfeld}}},
  \bibinfo{author}{\bibfnamefont{T.}~\bibnamefont{{Treu}}},
  \bibinfo{author}{\bibfnamefont{R.}~\bibnamefont{{Gavazzi}}},
  \bibinfo{author}{\bibfnamefont{S.~H.} \bibnamefont{{Suyu}}},
  \bibinfo{author}{\bibfnamefont{P.~J.} \bibnamefont{{Marshall}}},
  \bibinfo{author}{\bibfnamefont{M.~W.} \bibnamefont{{Auger}}},
  \bibnamefont{and} \bibinfo{author}{\bibfnamefont{C.}~\bibnamefont{{Nipoti}}},
  \bibinfo{journal}{\apj} \textbf{\bibinfo{volume}{777}}, \bibinfo{eid}{98}
  (\bibinfo{year}{2013}), \eprint{1307.4759}.

\bibitem[{\citenamefont{{Cao} et~al.}(2016)\citenamefont{{Cao}, {Biesiada},
  {Yao}, and {Zhu}}}]{holanda2017}
\bibinfo{author}{\bibfnamefont{S.}~\bibnamefont{{Cao}}},
  \bibinfo{author}{\bibfnamefont{M.}~\bibnamefont{{Biesiada}}},
  \bibinfo{author}{\bibfnamefont{M.}~\bibnamefont{{Yao}}}, \bibnamefont{and}
  \bibinfo{author}{\bibfnamefont{Z.-H.} \bibnamefont{{Zhu}}},
  \bibinfo{journal}{Mont. Not. Royal. Astron. Soc.}
  \textbf{\bibinfo{volume}{461}}, \bibinfo{pages}{2192} (\bibinfo{year}{2016}),
  \eprint{1604.05625}.

\bibitem[{\citenamefont{{Chen} et~al.}(2019)\citenamefont{{Chen}, {Li}, {Shu},
  and {Cao}}}]{YCHEN}
\bibinfo{author}{\bibfnamefont{Y.}~\bibnamefont{{Chen}}},
  \bibinfo{author}{\bibfnamefont{R.}~\bibnamefont{{Li}}},
  \bibinfo{author}{\bibfnamefont{Y.}~\bibnamefont{{Shu}}}, \bibnamefont{and}
  \bibinfo{author}{\bibfnamefont{X.}~\bibnamefont{{Cao}}},
  \bibinfo{journal}{Mont. Not. Royal. Astron. Soc.}
  \textbf{\bibinfo{volume}{488}}, \bibinfo{pages}{3745} (\bibinfo{year}{2019}),
  \eprint{1809.09845}.

\bibitem[{\citenamefont{{Minazzoli} and {Hees}}(2014)}]{hees2}
\bibinfo{author}{\bibfnamefont{O.}~\bibnamefont{{Minazzoli}}} \bibnamefont{and}
  \bibinfo{author}{\bibfnamefont{A.}~\bibnamefont{{Hees}}},
  \bibinfo{journal}{\prd} \textbf{\bibinfo{volume}{90}}, \bibinfo{eid}{023017}
  (\bibinfo{year}{2014}), \eprint{1404.4266}.

\bibitem[{\citenamefont{{Uzan}}(2011)}]{Uzan:2010pm}
\bibinfo{author}{\bibfnamefont{J.-P.} \bibnamefont{{Uzan}}},
  \bibinfo{journal}{Living Reviews in Relativity}
  \textbf{\bibinfo{volume}{14}}, \bibinfo{eid}{2} (\bibinfo{year}{2011}),
  \eprint{1009.5514}.

\bibitem[{\citenamefont{{Hees} et~al.}(2015)\citenamefont{{Hees}, {Minazzoli},
  and {Larena}}}]{Observables}
\bibinfo{author}{\bibfnamefont{A.}~\bibnamefont{{Hees}}},
  \bibinfo{author}{\bibfnamefont{O.}~\bibnamefont{{Minazzoli}}},
  \bibnamefont{and} \bibinfo{author}{\bibfnamefont{J.}~\bibnamefont{{Larena}}},
  \bibinfo{journal}{General Relativity and Gravitation}
  \textbf{\bibinfo{volume}{47}}, \bibinfo{eid}{9} (\bibinfo{year}{2015}),
  \eprint{1409.7273}.

\bibitem[{\citenamefont{{Damour}
  et~al.}(2002{\natexlab{b}})\citenamefont{{Damour}, {Piazza}, and
  {Veneziano}}}]{Damour12}
\bibinfo{author}{\bibfnamefont{T.}~\bibnamefont{{Damour}}},
  \bibinfo{author}{\bibfnamefont{F.}~\bibnamefont{{Piazza}}}, \bibnamefont{and}
  \bibinfo{author}{\bibfnamefont{G.}~\bibnamefont{{Veneziano}}},
  \bibinfo{journal}{\prl} \textbf{\bibinfo{volume}{89}}, \bibinfo{eid}{081601}
  (\bibinfo{year}{2002}{\natexlab{b}}), \eprint{gr-qc/0204094}.

\bibitem[{\citenamefont{Damour and Polyakov}(1994)}]{Damour:1994zq}
\bibinfo{author}{\bibfnamefont{T.}~\bibnamefont{Damour}} \bibnamefont{and}
  \bibinfo{author}{\bibfnamefont{A.~M.} \bibnamefont{Polyakov}},
  \bibinfo{journal}{Nucl. Phys. B} \textbf{\bibinfo{volume}{423}},
  \bibinfo{pages}{532} (\bibinfo{year}{1994}), \eprint{hep-th/9401069}.

\bibitem[{\citenamefont{{Scolnic} et~al.}(2018)\citenamefont{{Scolnic},
  {Jones}, {Rest}, {Pan}, {Chornock}, {Foley}, {Huber}, {Kessler}, {Narayan},
  {Riess} et~al.}}]{pantheon}
\bibinfo{author}{\bibfnamefont{D.~M.} \bibnamefont{{Scolnic}}},
  \bibinfo{author}{\bibfnamefont{D.~O.} \bibnamefont{{Jones}}},
  \bibinfo{author}{\bibfnamefont{A.}~\bibnamefont{{Rest}}},
  \bibinfo{author}{\bibfnamefont{Y.~C.} \bibnamefont{{Pan}}},
  \bibinfo{author}{\bibfnamefont{R.}~\bibnamefont{{Chornock}}},
  \bibinfo{author}{\bibfnamefont{R.~J.} \bibnamefont{{Foley}}},
  \bibinfo{author}{\bibfnamefont{M.~E.} \bibnamefont{{Huber}}},
  \bibinfo{author}{\bibfnamefont{R.}~\bibnamefont{{Kessler}}},
  \bibinfo{author}{\bibfnamefont{G.}~\bibnamefont{{Narayan}}},
  \bibinfo{author}{\bibfnamefont{A.~G.} \bibnamefont{{Riess}}},
  \bibnamefont{et~al.}, \bibinfo{journal}{\apj} \textbf{\bibinfo{volume}{859}},
  \bibinfo{eid}{101} (\bibinfo{year}{2018}), \eprint{1710.00845}.

\bibitem[{\citenamefont{Brout et~al.}(2022)}]{Brout:2022vxf}
\bibinfo{author}{\bibfnamefont{D.}~\bibnamefont{Brout}} \bibnamefont{et~al.}
  (\bibinfo{year}{2022}), \eprint{2202.04077}.

\bibitem[{\citenamefont{Schmelling}(1995)}]{Schmelling_1995}
\bibinfo{author}{\bibfnamefont{M.}~\bibnamefont{Schmelling}},
  \bibinfo{journal}{Physica Scripta} \textbf{\bibinfo{volume}{51}},
  \bibinfo{pages}{676} (\bibinfo{year}{1995}),
  \urlprefix\url{https://dx.doi.org/10.1088/0031-8949/51/6/002}.

\bibitem[{\citenamefont{{Chiba} and {Kohri}}(2003)}]{Chiba}
\bibinfo{author}{\bibfnamefont{T.}~\bibnamefont{{Chiba}}} \bibnamefont{and}
  \bibinfo{author}{\bibfnamefont{K.}~\bibnamefont{{Kohri}}},
  \bibinfo{journal}{Progress of Theoretical Physics}
  \textbf{\bibinfo{volume}{110}}, \bibinfo{pages}{195} (\bibinfo{year}{2003}),
  \eprint{astro-ph/0306486}.

\bibitem[{\citenamefont{{Kraiselburd} et~al.}(2015)\citenamefont{{Kraiselburd},
  {Landau}, {Negrelli}, and {Garc{\'\i}a-Berro}}}]{Kraiselburd}
\bibinfo{author}{\bibfnamefont{L.}~\bibnamefont{{Kraiselburd}}},
  \bibinfo{author}{\bibfnamefont{S.~J.} \bibnamefont{{Landau}}},
  \bibinfo{author}{\bibfnamefont{C.}~\bibnamefont{{Negrelli}}},
  \bibnamefont{and}
  \bibinfo{author}{\bibfnamefont{E.}~\bibnamefont{{Garc{\'\i}a-Berro}}},
  \bibinfo{journal}{\apss} \textbf{\bibinfo{volume}{357}}, \bibinfo{eid}{4}
  (\bibinfo{year}{2015}), \eprint{1412.3418}.

\bibitem[{\citenamefont{D’Agostino and Nunes}(2019)}]{Rocco_Nunes_2019}
\bibinfo{author}{\bibfnamefont{R.}~\bibnamefont{D’Agostino}}
  \bibnamefont{and} \bibinfo{author}{\bibfnamefont{R.~C.} \bibnamefont{Nunes}},
  \bibinfo{journal}{Physical Review D} \textbf{\bibinfo{volume}{100}}
  (\bibinfo{year}{2019}), ISSN \bibinfo{issn}{2470-0029},
  \urlprefix\url{http://dx.doi.org/10.1103/PhysRevD.100.044041}.

\bibitem[{\citenamefont{Maggiore
  et~al.}(2020{\natexlab{b}})\citenamefont{Maggiore, Broeck, Bartolo, Belgacem,
  Bertacca, Bizouard, Branchesi, Clesse, Foffa, García-Bellido
  et~al.}}]{ET_2020}
\bibinfo{author}{\bibfnamefont{M.}~\bibnamefont{Maggiore}},
  \bibinfo{author}{\bibfnamefont{C.~V.~D.} \bibnamefont{Broeck}},
  \bibinfo{author}{\bibfnamefont{N.}~\bibnamefont{Bartolo}},
  \bibinfo{author}{\bibfnamefont{E.}~\bibnamefont{Belgacem}},
  \bibinfo{author}{\bibfnamefont{D.}~\bibnamefont{Bertacca}},
  \bibinfo{author}{\bibfnamefont{M.~A.} \bibnamefont{Bizouard}},
  \bibinfo{author}{\bibfnamefont{M.}~\bibnamefont{Branchesi}},
  \bibinfo{author}{\bibfnamefont{S.}~\bibnamefont{Clesse}},
  \bibinfo{author}{\bibfnamefont{S.}~\bibnamefont{Foffa}},
  \bibinfo{author}{\bibfnamefont{J.}~\bibnamefont{García-Bellido}},
  \bibnamefont{et~al.}, \bibinfo{journal}{Journal of Cosmology and
  Astroparticle Physics} \textbf{\bibinfo{volume}{2020}},
  \bibinfo{pages}{050–050} (\bibinfo{year}{2020}{\natexlab{b}}), ISSN
  \bibinfo{issn}{1475-7516},
  \urlprefix\url{http://dx.doi.org/10.1088/1475-7516/2020/03/050}.

\bibitem[{\citenamefont{Zhao et~al.}(2011)\citenamefont{Zhao, Van Den~Broeck,
  Baskaran, and Li}}]{ET_2011}
\bibinfo{author}{\bibfnamefont{W.}~\bibnamefont{Zhao}},
  \bibinfo{author}{\bibfnamefont{C.}~\bibnamefont{Van Den~Broeck}},
  \bibinfo{author}{\bibfnamefont{D.}~\bibnamefont{Baskaran}}, \bibnamefont{and}
  \bibinfo{author}{\bibfnamefont{T.~G.~F.} \bibnamefont{Li}},
  \bibinfo{journal}{Physical Review D} \textbf{\bibinfo{volume}{83}}
  (\bibinfo{year}{2011}), ISSN \bibinfo{issn}{1550-2368},
  \urlprefix\url{http://dx.doi.org/10.1103/PhysRevD.83.023005}.

\bibitem[{\citenamefont{Cai and Yang}(2017{\natexlab{b}})}]{ET_2017}
\bibinfo{author}{\bibfnamefont{R.-G.} \bibnamefont{Cai}} \bibnamefont{and}
  \bibinfo{author}{\bibfnamefont{T.}~\bibnamefont{Yang}},
  \bibinfo{journal}{Physical Review D} \textbf{\bibinfo{volume}{95}}
  (\bibinfo{year}{2017}{\natexlab{b}}), ISSN \bibinfo{issn}{2470-0029},
  \urlprefix\url{http://dx.doi.org/10.1103/PhysRevD.95.044024}.

\bibitem[{\citenamefont{et~al}(2017)}]{amaroseoane2017laser}
\bibinfo{author}{\bibfnamefont{P.~A.-S.} \bibnamefont{et~al}},
  \emph{\bibinfo{title}{Laser interferometer space antenna}}
  (\bibinfo{year}{2017}), \eprint{1702.00786}.

\bibitem[{\citenamefont{Tamanini et~al.}(2016)\citenamefont{Tamanini, Caprini,
  Barausse, Sesana, Klein, and Petiteau}}]{Tamanini2016}
\bibinfo{author}{\bibfnamefont{N.}~\bibnamefont{Tamanini}},
  \bibinfo{author}{\bibfnamefont{C.}~\bibnamefont{Caprini}},
  \bibinfo{author}{\bibfnamefont{E.}~\bibnamefont{Barausse}},
  \bibinfo{author}{\bibfnamefont{A.}~\bibnamefont{Sesana}},
  \bibinfo{author}{\bibfnamefont{A.}~\bibnamefont{Klein}}, \bibnamefont{and}
  \bibinfo{author}{\bibfnamefont{A.}~\bibnamefont{Petiteau}},
  \bibinfo{journal}{Journal of Cosmology and Astroparticle Physics}
  \textbf{\bibinfo{volume}{2016}}, \bibinfo{pages}{002–002}
  (\bibinfo{year}{2016}), ISSN \bibinfo{issn}{1475-7516},
  \urlprefix\url{http://dx.doi.org/10.1088/1475-7516/2016/04/002}.

\bibitem[{\citenamefont{{Shu} et~al.}(2017)\citenamefont{{Shu}, {Brownstein},
  {Bolton}, {Koopmans}, {Treu}, {Montero-Dorta}, {Auger}, {Czoske}, {Gavazzi},
  {Marshall} et~al.}}]{Shu2017}
\bibinfo{author}{\bibfnamefont{Y.}~\bibnamefont{{Shu}}},
  \bibinfo{author}{\bibfnamefont{J.~R.} \bibnamefont{{Brownstein}}},
  \bibinfo{author}{\bibfnamefont{A.~S.} \bibnamefont{{Bolton}}},
  \bibinfo{author}{\bibfnamefont{L.~V.~E.} \bibnamefont{{Koopmans}}},
  \bibinfo{author}{\bibfnamefont{T.}~\bibnamefont{{Treu}}},
  \bibinfo{author}{\bibfnamefont{A.~D.} \bibnamefont{{Montero-Dorta}}},
  \bibinfo{author}{\bibfnamefont{M.~W.} \bibnamefont{{Auger}}},
  \bibinfo{author}{\bibfnamefont{O.}~\bibnamefont{{Czoske}}},
  \bibinfo{author}{\bibfnamefont{R.}~\bibnamefont{{Gavazzi}}},
  \bibinfo{author}{\bibfnamefont{P.~J.} \bibnamefont{{Marshall}}},
  \bibnamefont{et~al.}, \bibinfo{journal}{\apj} \textbf{\bibinfo{volume}{851}},
  \bibinfo{eid}{48} (\bibinfo{year}{2017}), \eprint{1711.00072}.

\bibitem[{\citenamefont{{Leaf} and {Melia}}(2018)}]{2018MNRAS.478.5104L}
\bibinfo{author}{\bibfnamefont{K.}~\bibnamefont{{Leaf}}} \bibnamefont{and}
  \bibinfo{author}{\bibfnamefont{F.}~\bibnamefont{{Melia}}},
  \bibinfo{journal}{Mont. Not. Royal. Astron. Soc.}
  \textbf{\bibinfo{volume}{478}}, \bibinfo{pages}{5104} (\bibinfo{year}{2018}),
  \eprint{1805.08640}.

\bibitem[{\citenamefont{{Foreman-Mackey}
  et~al.}(2013)\citenamefont{{Foreman-Mackey}, {Hogg}, {Lang}, and
  {Goodman}}}]{Foreman}
\bibinfo{author}{\bibfnamefont{D.}~\bibnamefont{{Foreman-Mackey}}},
  \bibinfo{author}{\bibfnamefont{D.~W.} \bibnamefont{{Hogg}}},
  \bibinfo{author}{\bibfnamefont{D.}~\bibnamefont{{Lang}}}, \bibnamefont{and}
  \bibinfo{author}{\bibfnamefont{J.}~\bibnamefont{{Goodman}}},
  \bibinfo{journal}{\pasp} \textbf{\bibinfo{volume}{125}}, \bibinfo{pages}{306}
  (\bibinfo{year}{2013}), \eprint{1202.3665}.

\bibitem[{\citenamefont{{Grillo}
  et~al.}(2008{\natexlab{a}})\citenamefont{{Grillo}, {Lombardi}, and
  {Bertin}}}]{Grillo}
\bibinfo{author}{\bibfnamefont{C.}~\bibnamefont{{Grillo}}},
  \bibinfo{author}{\bibfnamefont{M.}~\bibnamefont{{Lombardi}}},
  \bibnamefont{and} \bibinfo{author}{\bibfnamefont{G.}~\bibnamefont{{Bertin}}},
  \bibinfo{journal}{\aap} \textbf{\bibinfo{volume}{477}}, \bibinfo{pages}{397}
  (\bibinfo{year}{2008}{\natexlab{a}}), \eprint{0711.0882}.

\bibitem[{\citenamefont{{Grillo}
  et~al.}(2008{\natexlab{b}})\citenamefont{{Grillo}, {Lombardi}, and
  {Bertin}}}]{2008A&A...477..397G}
\bibinfo{author}{\bibfnamefont{C.}~\bibnamefont{{Grillo}}},
  \bibinfo{author}{\bibfnamefont{M.}~\bibnamefont{{Lombardi}}},
  \bibnamefont{and} \bibinfo{author}{\bibfnamefont{G.}~\bibnamefont{{Bertin}}},
  \bibinfo{journal}{\aap} \textbf{\bibinfo{volume}{477}}, \bibinfo{pages}{397}
  (\bibinfo{year}{2008}{\natexlab{b}}), \eprint{0711.0882}.

\bibitem[{\citenamefont{{Jorgensen} et~al.}(1995)\citenamefont{{Jorgensen},
  {Franx}, and {Kjaergaard}}}]{Jorgensen}
\bibinfo{author}{\bibfnamefont{I.}~\bibnamefont{{Jorgensen}}},
  \bibinfo{author}{\bibfnamefont{M.}~\bibnamefont{{Franx}}}, \bibnamefont{and}
  \bibinfo{author}{\bibfnamefont{P.}~\bibnamefont{{Kjaergaard}}},
  \bibinfo{journal}{Mont. Not. Royal. Astron. Soc.}
  \textbf{\bibinfo{volume}{276}}, \bibinfo{pages}{1341} (\bibinfo{year}{1995}).

\bibitem[{\citenamefont{Gelman and Rubin}(1992)}]{Gelman}
\bibinfo{author}{\bibfnamefont{A.}~\bibnamefont{Gelman}} \bibnamefont{and}
  \bibinfo{author}{\bibfnamefont{D.~B.} \bibnamefont{Rubin}},
  \bibinfo{journal}{Statistical Science} \textbf{\bibinfo{volume}{7}},
  \bibinfo{pages}{457 } (\bibinfo{year}{1992}),
  \urlprefix\url{https://doi.org/10.1214/ss/1177011136}.

\bibitem[{\citenamefont{Sereno et~al.}(2010)\citenamefont{Sereno, Sesana,
  Bleuler, Jetzer, Volonteri, and Begelman}}]{Sereno_2010}
\bibinfo{author}{\bibfnamefont{M.}~\bibnamefont{Sereno}},
  \bibinfo{author}{\bibfnamefont{A.}~\bibnamefont{Sesana}},
  \bibinfo{author}{\bibfnamefont{A.}~\bibnamefont{Bleuler}},
  \bibinfo{author}{\bibfnamefont{P.}~\bibnamefont{Jetzer}},
  \bibinfo{author}{\bibfnamefont{M.}~\bibnamefont{Volonteri}},
  \bibnamefont{and} \bibinfo{author}{\bibfnamefont{M.~C.}
  \bibnamefont{Begelman}}, \bibinfo{journal}{Physical Review Letters}
  \textbf{\bibinfo{volume}{105}} (\bibinfo{year}{2010}),
  \urlprefix\url{https://doi.org/10.1103%2Fphysrevlett.105.251101}.

\bibitem[{\citenamefont{Sereno et~al.}(2011)\citenamefont{Sereno, Jetzer,
  Sesana, and Volonteri}}]{Sereno_2011}
\bibinfo{author}{\bibfnamefont{M.}~\bibnamefont{Sereno}},
  \bibinfo{author}{\bibfnamefont{P.}~\bibnamefont{Jetzer}},
  \bibinfo{author}{\bibfnamefont{A.}~\bibnamefont{Sesana}}, \bibnamefont{and}
  \bibinfo{author}{\bibfnamefont{M.}~\bibnamefont{Volonteri}},
  \bibinfo{journal}{Monthly Notices of the Royal Astronomical Society}
  \textbf{\bibinfo{volume}{415}}, \bibinfo{pages}{2773} (\bibinfo{year}{2011}),
  \urlprefix\url{https://doi.org/10.1111%2Fj.1365-2966.2011.18895.x}.

\bibitem[{\citenamefont{Huang et~al.}(2023)\citenamefont{Huang, Hu, Chen, dong
  Zhang, Li, Gao, and Lin}}]{huang2023measuring}
\bibinfo{author}{\bibfnamefont{S.-J.} \bibnamefont{Huang}},
  \bibinfo{author}{\bibfnamefont{Y.-M.} \bibnamefont{Hu}},
  \bibinfo{author}{\bibfnamefont{X.}~\bibnamefont{Chen}},
  \bibinfo{author}{\bibfnamefont{J.}~\bibnamefont{dong Zhang}},
  \bibinfo{author}{\bibfnamefont{E.-K.} \bibnamefont{Li}},
  \bibinfo{author}{\bibfnamefont{Z.}~\bibnamefont{Gao}}, \bibnamefont{and}
  \bibinfo{author}{\bibfnamefont{X.-Y.} \bibnamefont{Lin}},
  \emph{\bibinfo{title}{Measuring the hubble constant using strongly lensed
  gravitational wave signals}} (\bibinfo{year}{2023}), \eprint{2304.10435}.

\bibitem[{\citenamefont{Liu et~al.}(2019)\citenamefont{Liu, Li, and
  Zhu}}]{Liu_2019}
\bibinfo{author}{\bibfnamefont{B.}~\bibnamefont{Liu}},
  \bibinfo{author}{\bibfnamefont{Z.}~\bibnamefont{Li}}, \bibnamefont{and}
  \bibinfo{author}{\bibfnamefont{Z.-H.} \bibnamefont{Zhu}},
  \bibinfo{journal}{Monthly Notices of the Royal Astronomical Society}
  \textbf{\bibinfo{volume}{487}}, \bibinfo{pages}{1980} (\bibinfo{year}{2019}),
  \urlprefix\url{https://doi.org/10.1093%2Fmnras%2Fstz1179}.

\bibitem[{\citenamefont{Hou et~al.}(2020)\citenamefont{Hou, Fan, Liao, and
  Zhu}}]{Hou_2020}
\bibinfo{author}{\bibfnamefont{S.}~\bibnamefont{Hou}},
  \bibinfo{author}{\bibfnamefont{X.-L.} \bibnamefont{Fan}},
  \bibinfo{author}{\bibfnamefont{K.}~\bibnamefont{Liao}}, \bibnamefont{and}
  \bibinfo{author}{\bibfnamefont{Z.-H.} \bibnamefont{Zhu}},
  \bibinfo{journal}{Physical Review D} \textbf{\bibinfo{volume}{101}}
  (\bibinfo{year}{2020}),
  \urlprefix\url{https://doi.org/10.1103%2Fphysrevd.101.064011}.

\bibitem[{\citenamefont{yi~Lin et~al.}(2023)\citenamefont{yi~Lin, dong Zhang,
  Dai, Huang, and Mei}}]{lin2023strong}
\bibinfo{author}{\bibfnamefont{X.}~\bibnamefont{yi~Lin}},
  \bibinfo{author}{\bibfnamefont{J.}~\bibnamefont{dong Zhang}},
  \bibinfo{author}{\bibfnamefont{L.}~\bibnamefont{Dai}},
  \bibinfo{author}{\bibfnamefont{S.-J.} \bibnamefont{Huang}}, \bibnamefont{and}
  \bibinfo{author}{\bibfnamefont{J.}~\bibnamefont{Mei}},
  \emph{\bibinfo{title}{Strong gravitational lensing of gravitational waves
  with tianqin}} (\bibinfo{year}{2023}), \eprint{2304.04800}.

\bibitem[{\citenamefont{Ali et~al.}(2023)\citenamefont{Ali, Stoikos, Meade,
  Kesden, and King}}]{Ali_2023}
\bibinfo{author}{\bibfnamefont{S.}~\bibnamefont{Ali}},
  \bibinfo{author}{\bibfnamefont{E.}~\bibnamefont{Stoikos}},
  \bibinfo{author}{\bibfnamefont{E.}~\bibnamefont{Meade}},
  \bibinfo{author}{\bibfnamefont{M.}~\bibnamefont{Kesden}}, \bibnamefont{and}
  \bibinfo{author}{\bibfnamefont{L.}~\bibnamefont{King}},
  \bibinfo{journal}{Physical Review D} \textbf{\bibinfo{volume}{107}}
  (\bibinfo{year}{2023}),
  \urlprefix\url{https://doi.org/10.1103%2Fphysrevd.107.103023}.

\end{thebibliography}
\label{lastpage}
\end{document}